\documentclass[12pt]{article}

\usepackage{euscript,amsmath, amssymb, amsfonts}

\newcommand{\supp}{{\rm supp\,}}
\newcommand{\R}{{\mathbb R}}

\newcommand{\oC}{{\mathbb C}}
\newcommand{\oP}{{\cal P}}
\newcommand{\di}{{\rm d}}
\newcommand{\D}{\EuScript D}
\newcommand{\eE}{\EuScript E}
\newcommand{\DD}{\overleftarrow D}
\newcommand{\G}{{\mathbb G}}
\newcommand{\Z}{{\mathbb Z}}
\newcommand{\K}{{\mathbb K}}
\newcommand{\be}{\begin{equation}}
\newcommand{\ee}{\end{equation}}

\newcounter{theorem}
\newcommand{\theorem}{\par\refstepcounter{theorem}
           {\bf Theorem \arabic{section}.\arabic{theorem}. }}
\renewcommand\thetheorem{\thesection.\arabic{theorem}}
\makeatletter \@addtoreset{theorem}{section}

\newcounter{lemma}
\newcommand{\lemma}{\par\refstepcounter{lemma}
           {\bf Lemma \arabic{section}.\arabic{lemma}. }}
\renewcommand\thelemma{\thesection.\arabic{lemma}}
\makeatletter \@addtoreset{lemma}{section}

\newcounter{proposition}
\newcommand{\proposition}{\par\refstepcounter{proposition}
           {\bf Proposition \arabic{section}.\arabic{proposition}. }}
\renewcommand\theproposition{\thesection.\arabic{proposition}}
\makeatletter \@addtoreset{proposition}{section}

\newcounter{appen}
\newcommand{\appen}{\par\refstepcounter{appen}
{\par\medskip\noindent\Large\bf Appendix \arabic{appen}\par\medskip }}

\font\frtnfr=eufm10   scaled\magstep1 \font\twlfr=eufm10
\font\tenfr=eufm10  
\newfam\frfam
\textfont\frfam=\frtnfr \scriptfont\frfam=\twlfr
\scriptscriptfont\frfam=\tenfr

\font\frtnopen=msbm10  scaled\magstep2 \font\twlopen=msbm10
\font\tenopen=msbm10  
\newfam\openfam
\textfont\openfam=\frtnopen \scriptfont\openfam=\twlopen
\scriptscriptfont\openfam=\tenopen
\def\open{\fam\openfam}

\font\frtnsf = cmss12 scaled\magstep1 \font\twlsf = cmss10
\font\tensf = cmss9  
\newfam\Scfam
\textfont\Scfam = \frtnsf \scriptfont\Scfam = \twlsf
\scriptscriptfont\Scfam = \tensf

\makeatletter \@addtoreset{equation}{section}
\def\theequation{\thesection.\arabic{equation}}
\makeatother

\begin{document}
\bibliographystyle{nphys}

\sloppy
\title
 {
           Cohomologies of the Poisson superalgebra
 }
\author
 {
 S.E.~Konstein\thanks{E-mail: konstein@lpi.ru}~,~A.G.~Smirnov{\thanks{E-mail:
 smirnov@lpi.ru}}~,~and~I.V.~Tyutin\thanks{E-mail: tyutin@lpi.ru}
% \thanks{
%               This work was supported
%               by the RFBR (grants No.~02-01-00930
%               (I.T.), No.~02-02-17067 (S.K.), and No.~02-02-16946 (A.S.)),
%               by INTAS grant No.~03-51-6346 (A.S.) and by the grant LSS-1578.2003.2.
% } 
\\
               {\sf \small I.E.Tamm Department of
               Theoretical Physics,} \\ {\sf \small P. N. Lebedev Physical
               Institute,} \\ {\sf \small 119991, Leninsky Prospect 53,
               Moscow, Russia.} } \date { }

\maketitle

\begin{abstract}
{ \footnotesize Cohomology spaces of the Poisson
superalgebra realized on smooth Grassmann-valued functions with compact
support on $\R^{2n}$ are investigated under suitable continuity restrictions
on cochains. The first and second cohomology spaces in the
trivial representation
and the zeroth and first cohomology spaces in the adjoint
representation of the Poisson superalgebra are found for the case of a
constant nondegenerate Poisson superbracket for arbitrary $n>0$. The third
cohomology space in the trivial representation and the second cohomology space
in the adjoint representation of this superalgebra are
found for arbitrary $n>1$.}
\end{abstract}

\section{Introduction}
\label{intr}

The hope to construct the quantum mechanics on nontrivial manifolds is
connected with geometrical or deformation quantization \cite{1} - \cite{4}.
The functions on the phase space are associated with the operators, and the
product and the commutator of the operators are described by associative
*-product and *-commutator of the functions. These *-product and *-commutator
are the deformations of usual product and of usual Poisson bracket.

To find the deformations of Poisson superalgebra, one should calculate the
second cohomology of the Poisson superalgebra.

In \cite{Zh}, the lower cohomologies (up to second) were calculated for the
Poisson algebra consisting of smooth complex-valued functions on ${\open
R}^{2n}$. In \cite{Leites}, the deformations of the Lie superalgebra of Hamiltonian
vector fields with polynomial coefficients are investigated.  The pure
Grassmanian case $n=0$ is considered in \cite{Ty1} and \cite{Ty2}.

In this paper, we calculate the lower cohomologies of the Poisson
superalgebra of the Grassmann-valued functions with compact supports
on ${\open R}^{2n}$ in the trivial (up to third cohomology) and adjoint
representation (up to second cohomology) for the case $n \geq 2$. It occurred
that the case $n=1$ needs a separate consideration which is performed in
\cite{n=2}. In \cite{central} we consider the central extensions of the
algebras considered in this paper.

Let $\K$ be either $\R$ or $\oC$. We denote by $\EuScript D(\R^n)$ the space
of smooth $\K$-valued functions with compact support on $\R^n$. This space is
endowed with its standard topology: by definition, a sequence $\varphi_k\in
\EuScript D(\R^n)$ converges to $\varphi\in \EuScript D(\R^n)$ if the
supports of all $\varphi_k$ are contained in a fixed compact set, and
$\partial^\lambda\varphi_k$ converge uniformly to $\partial^\lambda\varphi$
for every multi-index $\lambda$. We set $$ \mathbf D^{n_-}_{n_+}= \EuScript
D(\R^{n_+})\otimes \G^{n_-},\quad \mathbf E^{n_-}_{n_+}=
C^\infty(\R^{n_+})\otimes \G^{n_-},\quad \mathbf D^{\prime n_-}_{n_+}=
\EuScript D'(\R^{n_+})\otimes \G^{n_-}, $$ where $\G^{n_-}$ is the Grassmann
algebra with $n_-$ generators and $\EuScript D'(\R^{n_+})$ is the space of
continuous linear functionals on $\EuScript D(\R^{n_+})$. The generators of
the Grassmann algebra (resp., the coordinates of the space $\R^{n_+}$) are
denoted by $\xi^\alpha$, $\alpha=1,\ldots,n_-$ (resp., $x^i$, $i=1,\ldots,
n_+$). We shall also use collective variables $z^A$ which are equal to $x^A$
for $A=1,\ldots,n_+$ and are equal to $\xi^{A-n_+}$ for
$A=n_++1,\ldots,n_++n_-$. The spaces $\mathbf D^{n_-}_{n_+}$, $\mathbf
E^{n_-}_{n_+}$, and $\mathbf D^{\prime n_-}_{n_+}$ possess a natural grading
which is determined by that of the Grassmann algebra. The parity of an
element $f$ of these spaces is denoted by $\varepsilon(f)$. We also set
$\varepsilon_A=0$ for $A=1,\ldots, n_+$ and $\varepsilon_A=1$ for
$A=n_++1,\ldots, n_++n_-$.  Let
$\sigma(f,g)=(-1)^{\varepsilon(f)\varepsilon(g)}$ and
$\sigma(A,B)=(-1)^{\varepsilon_A\varepsilon_B}$.

Let $\partial/\partial z^A$ and $\overleftarrow{\partial}/\partial z^A$ be
the operators of the left and right differentiation. The Poisson bracket is
defined by the relation \begin{equation}
\{f,g\}(z)=f(z)\frac{\overleftarrow{\partial}}{\partial z^A}\omega^{AB}
\frac{\partial}{\partial z^B}g(z)= -\sigma(f,g)\{g,f\}(z),\label{3.0}
\end{equation} where the symplectic metric
$\omega^{AB}=-\sigma(A,B)\omega^{BA}$ is a constant invertible matrix.  For
 definiteness, we choose it in the form \[ \omega^{AB}=
\left(\begin{array}{cc}\omega^{ij}&0       \\
0&\lambda_\alpha\delta^{\alpha\beta}\end{array}\right),\quad
\lambda_\alpha=\pm1,\ i,j=1,...,n_+,\ \alpha,\beta=1+n_+,...,n_-+n_+, \]
where $\omega^{ij}$ is the canonical symplectic form (if $\K=\oC$, then one
can choose $\lambda_\alpha=1$). The nondegeneracy of the matrix $\omega^{AB}$
implies, in particular, that $n_+$ is even.  The Poisson superbracket
satisfies the Jacobi identity \begin{equation}
\sigma(f,h)\{f,\{g,h\}\}(z)+\hbox{cycle}(f,g,h)= 0,\quad f,g,h\in \mathbf
E^{n_-}_{n_+}. \label{3.0a} \end{equation} By Poisson superalgebra, we mean
the space $\mathbf D^{n_-}_{n_+}$ with the Poisson bracket~(\ref{3.0}) on it.
The relations~(\ref{3.0}) and~(\ref{3.0a}) show that this bracket indeed
determines a Lie superalgebra structure on $\mathbf D^{n_-}_{n_+}$.

The integral on $\mathbf D^{n_-}_{n_+}$ is defined by the relation $ \int \di
z\, f(z)= \int_{\R^{n_+}}\di x\int \di\xi\, f(z), $ where the integral on the
Grassmann algebra is normed by the condition $\int \di\xi\,
\xi^1\ldots\xi^{n_-}=1$. We identify $\G^{n_-}$ with its dual space
$\G^{\prime n_-}$ setting $f(g)=\int\di\xi\, f(\xi)g(\xi)$, $f,g\in
\G^{n_-}$.  Correspondingly, the space $\mathbf D^{\prime n_-}_{n_+}$ of
continuous linear functionals on $\mathbf D^{n_-}_{n_+}$ is identified  with
the space $\D^\prime(\R^{n_+})\otimes \G^{n_-}$.  As a rule, the value $m(f)$
of a functional $m\in \mathbf D^{\prime n_-}_{n_+}$ on a test function $f\in
\mathbf D^{n_-}_{n_+}$ will be written in the ``integral'' form:  $ m(f)
=\int \di z\, m(z) f(z).  $

Let $L$ be a Lie superalgebra acting in a $\Z_2$-graded space $V$ (the action
of $f\in L$ on $v\in V$ will be denoted by $f\cdot v$).  The space $C_p(L,
V)$ of $p$-cochains consists of all multilinear super-antisymmetric mappings
from $L^p$ to $V$. The space $C_p(L, V)$ possesses a natural $\Z_2$-grading:
by definition, $M_p\in C_p(L, V)$ has the definite parity $\varepsilon(M_p)$
if $$ \varepsilon(M_p(f_1,\ldots,f_p))=
\varepsilon(M_p)+\varepsilon(f_1)+\ldots+\varepsilon(f_p) $$ for any $f_j\in
L$ with parities $\epsilon(f_j)$. The differential $\di_p^V$ is defined to be
the linear operator from $C_p(L, V)$ to $C_{p+1}(L, V)$ such that
\begin{align} &&d_p^{V}M_p(f_1,...,f_{p+1})=
-\sum_{j=1}^{p+1}(-1)^{j+\varepsilon(f_j)|\varepsilon(f)|_{1,j-1}+
\varepsilon(f_j)\varepsilon_{M_p}}f_j\cdot
M_p(f_{1},...,\breve{f}_j,...,f_{p+1})- \nonumber \\
&&-\sum_{i<j}(-1)^{j+\varepsilon(f_j)|\varepsilon(f)|_{i+1,j-1}}
M_p(f_1,...f_{i-1},\{f_i,f_j\},f_{i+1},...,
\breve{f}_j,...,f_{p+1}),\label{diff} \end{align} for any $M_p\in C_p(L, V)$
and $f_1,\ldots f_{p+1}\in L$ having definite parities. Here the sign
$\breve{}$ means that the argument is omitted and the notation $$
|\varepsilon(f)|_{i,j}=\sum_{l=i}^j\varepsilon(f_l) $$ has been used. The
differential $\di^V$ is nilpotent (see \cite{Schei97}), {\it i.e.},
$\di^V_{p+1}\di^V_p=0$ for any $p=0,1,\ldots$. The $p$-th cohomology space of
the differential $\di_p^V$ will be denoted by $H^p_V$. The second cohomology
space $H^2_{\mathrm{ad}}$ in the adjoint representation is closely related to
the problem of finding formal deformations of the Lie bracket
$\{\cdot,\cdot\}$ of the form $ \{f,g\}_*=\{f,g\}+\hbar^2 \{f,g\}_1+\ldots. $
The condition that $\{\cdot,\cdot\}$ is a 2-cocycle is equivalent to the
Jacobi identity for $\{\cdot,\cdot\}_*$ modulo the $\hbar^4$-order terms.

In the case of Poisson algebra, the problem of finding deformations can
hardly be solved in such a general setting, and additional restrictions on
cochains (apart from linearity and antisymmetry) are usually imposed. In some
papers on deformation quantization it is supposed that the kernels of $n$-th
order deformations $\{\cdot,\cdot\}_n$ are bidifferential operators. In the
present paper, this requirement is replaced with the much weaker condition
that cochains should be separately continuous multilinear mappings.  It will
be shown that this gives rise to additional cohomologies.  
The general form of the deformations of the Poisson superalgebra considered
in the present paper is found in \cite{Ty3}.

We study the
cohomologies of the Poisson algebra $\mathbf D^{n_-}_{n_+}$ in the following
representations:
\begin{enumerate}
\item The trivial representation: $V=\K$,
$f\cdot a=0$ for any $f\in \mathbf D^{n_-}_{n_+}$ and $a\in\K$. The space
$C_p(\mathbf D^{n_-}_{n_+}, \K)$ consists of separately continuous
antisymmetric multilinear forms on $(\mathbf D^{n_-}_{n_+})^p$. The
cohomology spaces and the differentials will be denoted by
$H^p_{\mathrm{tr}}$ and $\di^{\mathrm{tr}}_p$ respectively.
\item $V=\mathbf
D^{\prime n_-}_{n_+}$ and $f\cdot m=\{f,m\}$ for any $f\in\mathbf
D^{n_-}_{n_+}$ and $m\in \mathbf D^{\prime n_-}_{n_+}$. The space
$C_p(\mathbf D^{n_-}_{n_+}, \mathbf D^{\prime n_-}_{n_+})$ consists of
separately continuous antisymmetric multilinear mappings from $(\mathbf
D^{n_-}_{n_+})^p$ to $\mathbf D^{\prime n_-}_{n_+}$. The continuity of $M\in
C_p(\mathbf D^{n_-}_{n_+}, \mathbf D^{\prime n_-}_{n_+})$ means that the
($p+1$)-form $$ (f_1,\ldots,f_{p+1})\to \int \di z\,
M(f_1,\ldots,f_p)(z)f_{p+1}(z) $$ on $(\mathbf D^{n_-}_{n_+})^{p+1}$ is
separately continuous. The cohomology spaces will be denoted by $H^p_{\mathbf
D'}$.
\item $V=\mathbf E^{n_-}_{n_+}$ and $f\cdot m=\{f,m\}$ for every
$f\in\mathbf D^{n_-}_{n_+}$ and $m\in \mathbf E^{n_-}_{n_+}$. The space
$C_p(\mathbf D^{n_-}_{n_+}, \mathbf E^{n_-}_{n_+})$ consists of separately
continuous antisymmetric multilinear mappings from $(\mathbf
D^{n_-}_{n_+})^p$ to $\mathbf E^{n_-}_{n_+}$ (the space $\mathbf
E^{n_-}_{n_+}$ is endowed with the topology of the uniform convergence on
compact subsets of $\R^{n_+}$). The cohomology spaces will be denoted by
$H^p_{\mathbf E}$.
\item The adjoint representation: $V=\mathbf
D^{n_-}_{n_+}$ and $f\cdot g=\{f,g\}$ for any $f,g\in\mathbf D^{n_-}_{n_+}$.
The space $C_p(\mathbf D^{n_-}_{n_+}, \mathbf D^{n_-}_{n_+})$ consists of
separately continuous antisymmetric multilinear mappings from $(\mathbf
D^{n_-}_{n_+})^p$ to $\mathbf D^{n_-}_{n_+}$. The cohomology spaces and the
differentials will be denoted by $H^p_{\mathrm{ad}}$ and
$\di^{\mathrm{ad}}_p$ respectively.  \end{enumerate} For the representations
2 and 3, we shall denote the differentials by the same symbol
$\di^{\mathrm{ad}}_p$ as in the adjoint representation. Note that in the case
of the trivial representation, $\K$ is considered as a graded space whose odd
subspace is trivial. We shall call p-cocycles $M_p^1,\ldots M_p^k$
independent if they give rise to linearly independent elements in $H^p$. For
a multilinear form $M_p$ taking values in $\mathbf D^{n_-}_{n_+}$, $\mathbf
E^{n_-}_{n_+}$, or $\mathbf D^{\prime n_-}_{n_+}$, we write
$M_p(z|f_1,\ldots,f_p)$ instead of more cumbersome
$M_p(f_1,\ldots,f_p)(z)$.

The main results of this work are given by the following three theorems.

%\textbf{Theorem 1.}
\theorem\label{th1}
\begin{enumerate}
\item $H^1_{\mathrm{tr}}\simeq \K$; the
linear form $M_1(f)=\bar f\stackrel{\mathrm{def}}{=}\int\di z\, f(z)$ is the
nontrivial cocycle.
\item Let bilinear forms $M_2^1$ and $M_2^2$ be defined
by the relations $$ M_2^1(f,g)=\bar f\bar g, \quad M_2^2(f,g)= \int
dz\left(f(z){}\eE_z g(z)-\sigma(f,g) g(z){}\eE_z f(z)\right),\quad f,g\in
\mathbf D^{n_-}_{n_+}, $$ where\footnote{The operator ${\cal E}_z$
is a derivation of the Poisson superalgebra.}
$\eE_z \stackrel {def} = 1-\frac 1 2 z^A
\frac {\partial} {\partial z^A}\,\,\,$.  

If $n_-$ is even and $n_++4\ne n_-$, then $H^2_{\mathrm{tr}}=0$; if $n_++4=
n_-$, then $H^2_{\mathrm{tr}}\simeq \K$ and the form $M^2_2$ is a nontrivial
cocycle; if $n_-$ is odd, then $H^2_{\mathrm{tr}}\simeq \K$ and the form
$M^1_2$ is a nontrivial cocycle.

\item 
Let $n_+\geq 4.$
and trilinear forms $M_3^1$, $M_3^2$, $M_3^3$, and $M_3^4$ be defined
by the relations
\begin{align} && M_3^1(f,g,h)=\int dzf(z)
\left[g(z)\left(\frac{\overleftarrow{\partial}}{\partial z^A}\omega^{AB}
\frac{\partial}{\partial z^B}\right)^3h(z)\right], \nonumber \\ 
&&
M_3^2(f,g,h)=\int dz\{f(z),g(z)\}h(z),\quad M_3^3(f,g,h)=\bar f\bar g\bar
h,\nonumber \\
&& M_3^4(f,g,h)=
\bar{f}M^2_2(g,h)-
\sigma(f,g)\bar{g}M^2_2(f,h)+\sigma(f,h)\sigma(g,h)\bar{h}M^2_2(f,g).  \nonumber
\end{align}

If $n_-$ is even, $n_+\ne n_-$, and $n_++4\ne n_-$, then
$H^3_{\mathrm{tr}}\simeq \K^2$ and the forms $M^1_3$ and $M^2_3$ are
independent nontrivial cocycle; if $n_+=n_-$, then $H^3_{\mathrm{tr}}\simeq
\K$ and the form $M^1_3$ is a nontrivial cocycle; if $n_++4=n_-$, then
$H^3_{\mathrm{tr}}\simeq \K^3$ and the forms $M^1_3$, $M^2_3$, and $M^4_3$
are independent nontrivial cocycles; if $n_-$ is odd, then
$H^3_{\mathrm{tr}}\simeq \K^3$ and the forms $M^1_3$, $M^2_3$, and $M^3_3$
are independent nontrivial cocycles.  \end{enumerate}

%\textbf{Theorem 2.}
\theorem\label{th2}
\begin{enumerate} 
\item $H^0_{\mathbf D'}\simeq
H^0_{\mathbf E}\simeq \K$; the function $M_0(z)\equiv 1$ is a nontrivial
cocycle.

\item $H^1_{\mathbf D'}\simeq H^1_{\mathbf E}\simeq \K^2$; independent
nontrivial cocycles are given by $$ M^1_1(z|f)=\bar f,\quad M^2_1(z|f)= \eE_z
f(z).  $$

\item Let $n_+\geq 4$ and the bilinear mappings $M_2^1$, $M_2^2$, $M_2^3$, and $M_2^4$ from
$(\mathbf D^{n_-}_{n_+})^2$ to $\mathbf E^{n_-}_{n_+}$ be defined by the
relations \begin{align} &&
M_2^1(z|f,g)=f(z)\!\left(\frac{\overleftarrow{\partial}}{\partial z^A}
\omega^{AB}\frac{\partial}{\partial z^B}\right)^3\!g(z), \quad
M^2_2(z|f,g)=\bar f \bar g,\nonumber \\ && M^3_2(z|f,g)= \bar g
M^2_1(z|f)-\sigma(f,g) \bar f M^2_1(z|g),\nonumber\\ && M^4_2(z|f,g)=\int
du\left(f(u) {}\eE_u g(u)- \sigma(f,g){g(u)}{}\eE_u f(u)\right).  \nonumber
\end{align} 

If $n_-$ is even and $n_++4\ne n_-$, then $H^2_{\mathbf D'}\simeq
H^2_{\mathbf E}\simeq \K^2$ and the cochains $M^1_2$ and $M^3_2$ are
independent nontrivial cocycles; if $n_++4= n_-$, then $H^2_{\mathbf
D'}\simeq H^2_{\mathbf E}\simeq \K^3$ and the cochains $M^1_2$, $M^3_2$, and
$M^4_2$ are independent nontrivial cocycles; if $n_-$ is odd, then
$H^2_{\mathbf D'}\simeq H^2_{\mathbf E}\simeq \K^3$ and the cochains $M^1_2$,
$M^2_2$ and $M^3_2$ are independent nontrivial cocycles.  \end{enumerate}

%\textbf{Theorem 3.}
\theorem\label{th3}
\begin{enumerate} \item $H^0_{\mathrm{ad}}=0$.

\item Let $V_1$ be the one-dimensional subspace of $C_1(\mathbf
D^{n_-}_{n_+},\mathbf D^{n_-}_{n_+})$ generated by the cocycle $M^2_1$
defined in Theorem~\ref{th2}. Then there is a natural isomorphism $V_1\oplus (\mathbf
E^{n_-}_{n_+}/\mathbf D^{n_-}_{n_+})\simeq H^1_{\mathrm{ad}}$ taking
$(M_1,T)\in V_1\oplus (\mathbf E^{n_-}_{n_+}/\mathbf D^{n_-}_{n_+})$ to the
cohomology class determined by the cocycle $M_1(z|f)+\{t(z),f(z)\}$, where
$t\in \mathbf E^{n_-}_{n_+}$ belongs to the equivalence class $T$.

\item Let $n_+\geq 4$. Let $V_2$ be the two-dimensional subspace of
$C_2(\mathbf D^{n_-}_{n_+},\mathbf D^{n_-}_{n_+})$ generated by the cocycles
$M^1_2$ and $M^3_2$ defined in Theorem~\ref{th2}. 
Then there is a natural isomorphism
$V_2\oplus (\mathbf E^{n_-}_{n_+}/\mathbf D^{n_-}_{n_+})\simeq
H^2_{\mathrm{ad}}$ taking $(M_2,T)\in V_2\oplus (\mathbf
E^{n_-}_{n_+}/\mathbf D^{n_-}_{n_+})$ to the cohomology class determined by
the cocycle $ M_2(z|f,g)-\{t(z),f(z)\}\bar{g}+
\sigma(f,g)\{t(z),g(z)\}\bar{f}, $ where $t\in \mathbf E^{n_-}_{n_+}$
belongs to the equivalence class $T$.  
\end{enumerate}

\section{Preliminaries} \label{prel} We adopt the following multi-index
notation:  \[ (z^A)^0\equiv1,\quad (z^A)^k\equiv z^{A_1}\cdots z^{A_k},\quad
k\geq 1, \] \[ (\partial^z_A)^0\equiv1,\quad (\partial^z_A)^k\equiv
\frac{\partial}{\partial z^{A_1}}\cdots \frac{\partial}{\partial
z^{A_k}},\quad k\geq 1, \] \[
(\overleftarrow{\partial}^z_A)^0\equiv1,\quad(\overleftarrow{\partial}^z_A)^k
\equiv \frac{\overleftarrow{\partial}}{\partial z^{A_k}}\cdots
\frac{\overleftarrow{\partial}}{\partial z^{A_1}},\quad k\geq 1, \] \[
T^{\ldots(A)_k\ldots }\equiv T^{\ldots A_1\ldots A_k\ldots },\quad T^{\ldots
A_iA_{i+1}\ldots }=\sigma(A_i,A_{i+1})T^{\ldots A_{i+1}A_i\ldots },\quad
i=1,\ldots,k-1.  \]

The $\delta$-function is normed by the relation 
$$ \int dz^{\prime }\delta
(z^{\prime }-z)f(z^{\prime })=\int f(z^{\prime })\delta (z-z^{\prime
})dz^{\prime }=f(z). $$

We denote by $M_p(\ldots)$ the separately continuous $p$-linear forms on
$(\mathbf D^{n_-}_{n_+})^p$. Thus, the arguments of these functionals are the
functions $f(z)$ of the form \begin{equation}\label{dec}
f(z)=\sum_{k=0}^{n_-} f_{(\alpha)_k}(x)(\xi^\alpha)^k,\quad
f_{(\alpha)_k}(x)\in \D(\R^{n_+}).  \end{equation} It can be easily proved
that such multilinear forms can be written in the integral form (see
Appendix~\ref{App2}):  
\begin{equation} M_p(f_1,\ldots,f_p)=\int dz_p\cdots
dz_1m_p(z_1,\ldots,z_p) f_1(z_1)\cdots f_p(z_p),\;p=1,2,... \label{3.1}
\end{equation} 
and 
\begin{equation} M_p(z|f_1,\ldots,f_p)=\int dz_p\cdots
dz_1m_p(z|z_1,\ldots,z_p) f_1(z_1)\cdots
f_p(z_p),\;p=1,2,\,...\,\,.\label{3.2} \end{equation} 
It follows from the
antisymmetry properties of the forms $M_p$ that the corresponding kernels
$m_p$ have the following properties:  
\begin{eqnarray}
\varepsilon(m_p(*|z_1,\ldots,z_p))=pn_{-}+\varepsilon_{M_p}, \nonumber
\\ 
m_p(*|z_1\ldots z_i,z_{i+1}\ldots z_p)=-(-1)^{n_-} m_p(*|z_{1}\ldots
z_{i+1},z_i\ldots z_p). \label{3.2a} 
\end{eqnarray} 
For the trivial and
adjoint representations of the Poisson superalgebra, the general
definition~(\ref{diff}) takes the form 
\begin{eqnarray} 
\!\!\! d_p^{\rm tr}M_{p}(f)
=-\sum_{i<j}(-1)^{j+\varepsilon(f_j)|\varepsilon(f)|_{i+1,j-1}}
M_p(f_1,...f_{i-1},\{f_i,f_j\},f_{i+1},...,\breve{f}_j,...,f_{p+1}),\ 
p \ge 1, \label{3.2b} \\ 
\!\!\! d_p^{\rm ad}M_p(z|f)
=\sum_{j=1}^{p+1}(-1)^{j+\varepsilon(f_j)|\varepsilon(f)|_{j+1,p+1}}
\{M_p(z|f_{1},...,\breve{f}_j,...,f_{p+1}),\, f_j(z)\}- \nonumber \\
\!\!\! -\sum_{i<j}(-1)^{j+\varepsilon(f_j)|\varepsilon(f)|_{i+1,j-1}}
M_p(z|f_1,...f_{i-1},\{f_i,f_j\},f_{i+1},...,\breve{f}_j,...,f_{p+1}),\ 
p=0,1,\ldots, \label{3.2c} 
\end{eqnarray} 
where the sign $\breve{}$ over the
argument means that this argument is omitted.

The support $\supp(f)\subset \R^{n_+}$ of a test function $f\in \mathbf
D^{n_-}_{n_+}$ is defined to be the union of the supports of all
$f_{(\alpha)_k}$ entering in the decomposition~(\ref{dec}).  For $m\in
\mathbf D^{\prime n_-}_{n_+}$, the set $\supp(m)$ is defined analogously.
Besides, we introduce the following notation \begin{eqnarray}
\DD_z^A\stackrel {def} = \frac {\overleftarrow {\partial}}{\partial
z^B}\omega^{BA}.  \end{eqnarray}

%@@@@@@@@@@@@@@@@@@@@@@@@@@@@@@@

The Poisson superbracket has the following useful property:
\begin{equation}
\int dzf(z)\{g(z),h(z)\}=\int dz\{f(z),g(z)\}h(z) \label{3.0aa}
\end{equation}
for arbitrary test functions $f,g,h$.

%@@@@@@@@@@@@@@@@@@@@@@@@@@@@@@@

\section{Cohomologies in the trivial representation} \label{triv}

In this section, we prove Theorem~\ref{th1} describing the first, second, and third
cohomologies in the trivial representation.

\subsection{$H^1_{\rm tr}$} \label{Htr1}

Let $M_1(f)=\int dzm_1(z)f(z)$. Then the cohomology equation has the form
\begin{equation} 
0=-d^{\rm tr}_1M_1(f,g)=M_1(\{f,g\})= 
\int dz\,m_1(z)\{f(z),g(z)\}=\int dz\{m_1(z),f(z)\}g(z). \label{4.1}
\end{equation} 
It follows from (\ref{4.1}) that
$$m_1(z)\frac{\overleftarrow{\partial}}{\partial z^A}\omega^{AB}
\frac{\partial}{\partial z^B}f(z)=0$$, 
which implies that
$$m_1(z)\frac{\overleftarrow{\partial}}{\partial z^A}=0 \ \mbox{ and }\  
m_1(z)=m_1={\rm
const}.$$

We therefore have $M_1(f)=m_1\bar{f}$,
$\,\,\varepsilon(m_1)=n_-+\varepsilon(M_1)$, where $ \bar{f}\equiv\int
dzf(z)$. Obviously, these cocycles are nontrivial.

\subsection{$H^2_{\rm tr}$} \label{Htr2}

{}For the bilinear form $M_2(f,g)=\int du\,dz\,m_2(z,u)f(z)g(u)$, the
cohomology equation has the form 
\begin{equation} 
0=-d^{\rm tr}_2 M_2(f,g,h) =
M_2(\{f,g\},h)-\sigma(g,h)M_2(\{f,h\},g)- M_2(f,\{g,h\}).
\label{4.2} \end{equation}

Let $ {\rm supp}(h)\bigcap\left[{\rm supp}(f)\bigcup{\rm
supp}(g)\right]=\varnothing. $ Then we have $ \hat{M}_2(\{f,g\},h)=0 $ or
\begin{equation} 
\int dz\,\hat{m}_2(z,u)\{f(z),g(z)\}=0\quad\Longrightarrow\quad
\frac{\partial\hat{m}_2(z,u)}{\partial z^A}=0, \label{4.2a} 
\end{equation}
where the hat means that the corresponding form (or kernel) is considered out
of the diagonal $z=u$.

Let $T_A(z,u)={\partial m_2(z,u)}/{\partial z^A}$. Then we have
$\hat{T}_A(z,u)=0$ and by Lemma~\ref{A5.2}, 
there is a locally finite decomposition
$$ T_A(z,u)=\sum_{q=0}^Q(\partial^z_C)^q\delta(z-u)t_A^{(C)_q}(u).  $$ When
working with decompositions of this type, we shall always first restrict them
to a bounded domain, where the summation limit $Q$ can be chosen finite, and
then ``glue'' together the results obtained for different domains. In every
particular case, the possibility of such gluing will be obvious. Everywhere
below the summation limit $Q$ is understood as a finite number depending
implicitly on a chosen bounded domain.

By construction, the kernel $T_A(z,u)$ has the property $ \partial^z_A
T_B(z,u)-\sigma(A,B) \partial^z_B T_A(z,u)=0.  $

This implies that $$ T_A(z,u)=\frac{\partial}{\partial
z^A}\sum_{q=0}^Q(\partial_C)^q\delta(z-u) t^{(C)_q}(u)$$ and $$
\frac{\partial}{\partial z^A}\left[m_2(z,u)-\sum_{q=0}^Q
(\partial^z_C)^q\delta(z-u)t^{(C)_q}(u) \right]=0.  $$ So 
$$
m_2(z,u)-\sum_{q=0}^Q(\partial^z_C)^q\delta(z-u)t^{(C)_q}(u)=m(u).$$  

By the
antisymmetry of $m_2(z,u)$, we have $m(u)=c= {\rm const},\quad
\varepsilon(c)=\varepsilon(M_2)$. We thus obtain
$M_2(f,g)=M_{2|1}(f,g)+M_{2|2}(f,g)$, where
$M_{2|1}(f,g)=c\bar{f}\bar{g},\;d_2^{\rm tr}M_{2|1}(f,g,h)=0$, and $M_{2|2}$
has the form \begin{equation} M_{2|2}(f,g)=\sum_{q=1}^Q\int
dz\,m_2^{(A)_q}(z)\left(
[(\partial^z_A)^qf(z)]g(z)-\sigma(f,g)[(\partial^z_A)^q g(z)]f(z)\right).
\label{4.3} \end{equation} Because of the antisymmetry of $M_2$, the constant
$c$ can be nonzero only for odd $n_-$.

Using integration by parts, we can rewrite the expression (\ref{4.3}) as a
sum over odd $q$ only and represent $M_{2|2}(f,g)$ in the form
\begin{equation} 
M_{2|2}(f,g)=\sum_{q=1}^{Q}\int
dzm_2^{(A)_q}(z)(\partial_A^u)^q[f(z+u)g(z-u)]|_{u=0}.
%\nonumber
\end{equation} 
The cohomology equation (\ref{4.2}) now yields
\begin{multline}\label{0000} \sum_{q=1}^{Q}\int
dz\,m_2^{(A)_q}(z)(\partial_A^u)^q [\{f,g\}(z+u)h(z-u)- \\
\sigma(g,h)\{f,h\}(z+u)g(z-u)-f(z+u)\{g,h\}(z-u)]|_{u=0}=0.  \end{multline}

%{\bf Proposition 3.2.1}
\proposition\label{q3.2.1}
$m_2^{(A)_q}(z)=0$ if $q>1$.

Indeed, let $Q\geq 3$. Substituting $f(z)=e^{zp_1}\varphi_1(z)$,
$g(z)=e^{zp_2}\varphi_2(z)$, and $h(z)=e^{-z(p_1+p_2)}\varphi_3(z)$ in the
left-hand side of this equation yields a polynomial in $p_1$ and $p_2$.
Equation~(\ref{0000}) implies that its homogeneous part of degree $Q+2$
should vanish:  \[ \int dz\,m_2^{(A)_Q}\varphi_1\varphi_2\varphi_3
[p_1\,,p_2] [((p_1+p_2)_A)^Q-(p_{2A})^Q-(p_{1A})^Q]=0 \] (here and below
$[p_1,p_2]=(-1)^{\varepsilon_B}p_{1B}\omega^{BC}p_{2C}$). This implies the
required statement.

Thus, $M_{2|2}$ has the form \[ M_{2|2}(f,g) =\int
dz\,m_2^A(z)\left(2\frac{\partial f(z)}{\partial z^A}g(z)+
(-1)^{\varepsilon_A}\frac{\overleftarrow{\partial}}{\partial z^A}
f(z)g(z)\right), \] and the equation for $m_2^A(z)$ takes the form
\begin{eqnarray} 
2\int dz\,m_2^A\left(\frac{\partial\{f,g\}}{\partial z^A}h-
\sigma(g,h)\frac{\partial\{f,h\}}{\partial z^A}g- \frac{\partial f}{\partial
z^A}\{g,h\}\right) &+& \nonumber \\ +\int
dz\,m\left(\{f,g\}h-\sigma(g,h)\{f,h\}g-f\{g,h\}\right) &=& 0, \nonumber
\end{eqnarray} 
where $$ m(z)=(-1)^{\varepsilon_{A}}m_2^A(z)
\frac{\overleftarrow{\partial}}{\partial z^A}.$$ Varying this equation w.r.t.
$h(z)$, we obtain 
\begin{eqnarray} 2[\sigma(A,f)\{m_2^A,f\}\frac{\partial
g}{\partial z^A}- \sigma(f,g)\sigma(A,g)\{m^A_2,g\}\frac{\partial f}{\partial
z^A}]-m\{f,g\} &+& \nonumber \\ + \{m,f\}g-\sigma(f,g)\{m,g\}f &=& 0.
\nonumber 
\end{eqnarray} 
Putting the coefficients at $f(z)$ and $\partial
f(z)\partial g(z)$ equal to zero, we obtain
$m(z){\overleftarrow{\partial}}/{\partial z^A}=0$. So $m(z)=-m={\rm
const}$, $\varepsilon(m)=n_-+\varepsilon(M_2)$, and the equation for
$m_2^A(z)$ assumes the form \begin{equation}
2m_2^A(z)\frac{\overleftarrow{\partial}}{\partial z^C}\omega^{CB}-
2\sigma(A,B)m_2^B(z)\frac{\overleftarrow{\partial}}{\partial z^C}\omega^{CA}-
m\omega ^{AB}=0. \label{4.4} \end{equation} Multiplying Eq. (\ref{4.4}) by
$(-1)^{\varepsilon_{A}}\omega _{BA}$ gives \begin{equation}
(4+n_{+}-n_{-})m=0. \label{4.4a} \end{equation}

To solve Eq. (\ref{4.4a}), we have to consider two cases.

i) Let $4+n_{+}-n_{-}\neq 0$. Then $m=0$ and by the Poincare lemma we have $
m_2^A(z)=\frac{1}{2}m_2(z) \DD_z^A ,\quad
\varepsilon(m_2)=n_-+\varepsilon_{M_2}. $

ii) Let $4+n_+-n_-=0$. In this case, the general solution of Eq. (\ref{4.4})
has the form \[ m_2^A(z)=\frac{1}{4}mz^A+ \frac{1}{2}m_2(z) \DD_z^A.  \]

{}Finally, we obtain 
\begin{eqnarray}
M_2(f,g)=\frac{1-(-1)^{n_-}}{2}c\bar{f}\bar{g} +\nonumber \\
+\delta_{n_--n_+,4} m\int \!\!
dz \left(f(z) \eE_z g(z)-\sigma(f,g) g(z) \eE_z  f(z)\right)+ d_1^{\rm
tr}m_1(f,g).  \label{4.5} 
\end{eqnarray}

The first two terms in (\ref{4.5}) are independent nontrivial cocycles.
Indeed, let 
\begin{equation} c\bar{f}\bar{g} +\delta_{n_--n_+,4} m\int \!\!
dz \left(f(z) \eE_z g(z)-\sigma(f,g) g(z) \eE_z  f(z)\right)= d_1^{\rm
tr}m_1(f,g)=-m_1(\{f,g\}). \label{4.6} 
\end{equation} 
Suppose 
that \[ {\rm supp}(f)\bigcap{\rm supp}(g)=\varnothing.  \] It follows
from (\ref{4.6}) that $ c\bar{f}\bar{g}=0$ and so $c=0$.

Let $g(z)=1$ in a neighborhood of ${\rm supp}(f)$. Then the Eq. (\ref{4.6})
is reduced to $$m\int dz\, z^A \partial_A f(z)=4m\bar{f}=0 \ 
\mbox{ and so }\  m=0,$$ {\it
i.e.} the sum of the first two terms is a coboundary only if $c=m=0$.

\subsection{$H^3_{\rm tr}$} \label{Htr3}

In this subsection we assume that $n_+\ge 4$.

The cohomology equation for the form
\[
M_3(f,g,h)=\int dvdudzm_3(z,u,v)f(z)g(u)h(v),
\]
is given by
\[
d_3^{\rm tr}M_3(f,g,h,s)=0=
\]
\[
=M_3(\{f,g\},h,s)-\sigma(g,h)M_3(\{f,h\},g,s)+
\sigma(g,s)\sigma(h,s)M_3(\{f,s\},g,h)-
\]
\begin{equation}
-M_3(f,\{g,h\},s)+\sigma(h,s)M_3(f,\{g,s\},h)+M_3(f,g,\{h,s\}). \label{4.7}
\end{equation}

Let the arguments $f$, $g$ and $h$ of the form $M_3$ be such that
\[
{\rm supp}(s)\bigcap\left[{\rm supp}(f)\bigcup{\rm supp}(g)\bigcup
{\rm supp}(h)\right]={\rm supp}(h)\bigcap\left[{\rm supp}(f)\bigcup
{\rm supp}(g)\right]=\varnothing.
\]
In this case, the cohomology equation (\ref{4.7}) is reduced  to
\[
\hat{M}_3(\{f,g\},h,s)=0\;\Longrightarrow
\]
\begin{equation}
\partial_A\hat{m}_3(z,u,v)=0=\partial_A^u\hat{m}_3(z,u,v)=
\partial_A^v\hat{m}_3(z,u,v). \label{4.8}
\end{equation}
The equation (\ref{4.8}) can be solved in the same manner as
(\ref{4.2a}). As a result, we have
\[
M_3=M_{3|1}+M_{3|2},
\]
\[
M_{3|1}(f,g,h)=c\bar{f}\bar{g}\bar{h},\quad
\varepsilon(c)=n_-+\varepsilon(M_3),
\]
\begin{eqnarray}
M_{3|2}(f,g,h)=\sum_{q=0}^Q\left[m^{(A)_q}([(\partial_A)^qf]g,h)-
\sigma(f,g)m^{(A)_q}([(\partial_A)^qg]f,h)\right]- \nonumber  \\
-\sigma(g,h)\sum_{q=0}^Q\left[m^{(A)_q}
([(\partial_A)^qf]h,g)-\sigma(f,h)m^{(A)_q}([(\partial_A)^qh]f,g)\right]+
\nonumber   \\
+\sigma(f,g)\sigma(f,h)\sum_{q=0}^Q\left[
m^{(A)_q}([(\partial_A)^qg]h,f)-\sigma(g,h)
m^{(A)_q}([(\partial_A)^qh]g,f)\right], \nonumber
\end{eqnarray}
\[
\varepsilon(m^{(A)_q})=|\varepsilon_A|_{1,q}+\varepsilon(M_3).
\]
Now let the functions in (\ref{4.7})
be such that
\[
{\rm supp}(s)\bigcap\left[{\rm supp}(f)\bigcup{\rm supp}(g)\bigcup
{\rm supp}(h)\right]=\varnothing.
\]
Then we have
\[
\hat{M}_2(f,g)=\sum_{q=1}^Q\int dudzm^{(A)_q}(z,u)\left([(\partial^z_A)^qf(z)]
g(z)-\sigma(f,g)[(\partial^z_A)^qg(z)]f(z)\right)s(u),
\]
where
$s(u)$
is a fixed function.
Acting as in Subsection~\ref{Htr2}
we obtain
\[
d_2^{\rm tr}\hat{M}_2(f,g,h)=0\;\Longrightarrow
\]
\[
\hat{m}^{(A)_q}(z,u)=0,\;q>1.
\]

As above, we consider cases $4+n_+-n_-\neq0$ and $4+n_+-n_-=0$ separately.

\subsubsection{The case
$\mathbf{4+n_+-n_-\neq0\,.}$}\label{i)}
\[
\hat{m}^A(z,u)\frac{\overleftarrow{\partial}}{\partial z^C}\omega^{CB}-
\sigma(A,B)\hat{m}^B(z,u)
\frac{\overleftarrow{\partial}}{\partial z^C}\omega^{CA}=0.
\]

The form
\[
T^{AB}(z,u)\equiv m^A(z,u)
\frac{\overleftarrow{\partial}}{\partial z^C}\omega^{CB}-\sigma(A,B)m^B(z,u)
\frac{\overleftarrow{\partial}}{\partial z^C}\omega^{CA}
\]
possesses the property
\[
\hat{T}^{AB}(z,u)=0
\]
and, therefore, can be written in the form
\[
T^{AB}(z,u)=\sum_{q=0}^Qt^{AB|(C)_q}(u)(\partial^u_C)^q\delta(u-z).
\]
From the definition of
$T^{AB}(z,u)$ it follows
that
\[
\sigma(A,C)T^{AB}(z,u)\frac{\overleftarrow{\partial}}{\partial z^D}\omega^{DC}
+{\rm cycle}(A,B,C)= 0\;\Longrightarrow
\]
\[
\sigma(A,C)\sum_{q=0}^Qt^{AB|(C)_q}(u)(\partial^u_{C})^q\delta(u-z)
\frac{\overleftarrow{\partial}}{\partial z^D}\omega^{DC}+
{\rm cycle}(A,B,C)= 0\;\Longrightarrow
\]
\[
\sum_{q=0}^Qt^{AB|(C)_q}(u)(\partial^u_C)^q\delta(u-z)=\sum_{q=0}^Q
t^{A|(C)_q}(u)(\partial^u_C)^q\delta(u-z)
\frac{\overleftarrow{\partial}}{\partial z^C}\omega^{CB}-
\sigma(A,B)(A\leftrightarrow B)\;\Longrightarrow
\]
\begin{equation}
\mu^A(z,u)\frac{\overleftarrow{\partial}}{\partial z^C}\omega^{CB}-
\sigma(A,B)\mu^B(z,u)\frac{\overleftarrow{\partial}}{\partial z^C}\omega^{CA}
=0, \label{4.9}
\end{equation}
where
\[
\mu^A(z,u)=m^A(z,u)-\sum_{q=0}^Qt^{A|(C)_q}(u)(\partial^u_C)^q\delta(u-z).
\]
Relation (\ref{4.9}) implies that
\[
m^A(z,u)=\frac{1}{2}m(z,u)\frac{\overleftarrow{\partial}}{\partial z^B}
\omega^{BA}+\sum_{q=0}t^{A|(C)_q}(u)(\partial^u_C)^q\delta(u-z).
\]
We thus have
\[
M_{3|2}(f,g,h)=m(\{f,g\},h)-\sigma(g,h)m(\{f,h\},g)+
\sigma(f,g)\sigma(f,h)m(\{g,h\},f)+M_{3|2\mathrm{loc}},
\]
\[
M_{3|2\mathrm{loc}}=\int dz\sum_{p,q,l=0}^Nm^{(A)_p|(B)_q|(C)_l}(z)
([(\partial^z_A)^pf(z)][(\partial^z_B)^qg(z)][(\partial^z_C)^lh(z)]).
\]
Substituting this representation for
$M_{3|2}$
in the equation
(\ref{4.7}) and choosing the functions $f$, $g$ and $h$ such that
\[
\left[{\rm supp}(f)\bigcup{\rm supp}(g)\right]\bigcap\left[
{\rm supp}(h)\bigcup{\rm supp}(s)\right]=\varnothing,
\]
we obtain
\[
\hat{m}(\{f,g\},\{h,s\})+\sigma(f,h)\sigma(f,s)\sigma(g,h)\sigma(g,s)
\hat{m}(\{h,s\},\quad\{f,g\})=0\,,
\]
so
\[
\int du\left[\{\hat{m}(z,u),g(z)\}\{h(u),s(u)\}+\right.
\]
\begin{equation}
\left.+(-1)^{n_-}\sigma(g,h)\sigma(g,s)
\{\hat{m}(u,z)\{h(u),s(u)\},g(z)\}\right]=0\,
\label{4.10}
\end{equation}
and therefore 
\[
\frac{\partial}{\partial z^B}\int du\left[\hat{m}(z,u)+
(-1)^{n_-}\hat{m}(u,z)\right]\{h(u),s(u)\}=0\,,
\]
which implies
\[
\frac{\partial}{\partial z^B}(\hat{m}(z,u)+(-1)^{n_{-}}\hat{m}(u,z))
\frac{\overleftarrow{\partial}}{\partial u^C}=0.
\]
Let
\[
T(z,u)\equiv m(z,u)+(-1)^{n_-}m(u,z),\quad
T_B(z,u)\equiv\frac{\partial}{\partial z^B}T(z,u),
\]
\[
T_{BC}(z,u)\equiv T_B(z,u)\frac{\overleftarrow{\partial}}{\partial u^C}.
\]
Since $\hat{T}_{BC}(z,u)=0$, we can represent the distribution ${T}_{BC}(z,u)$ in the form
\[
T_{BC}(z,u)=\sum_{q=0}^Qt_{BC}^{(A)_q}(z)(\partial^z_A)^q\delta(z-u).
\]

By the construction of the distribution
$T_{BC}(z,u)$, we have
\[
T_{BC}(z,u)\frac{\overleftarrow{\partial}}{\partial u^D}-
\sigma(C,D)(C\leftrightarrow D)=0\,.
\]
So
\[
t_{BC}^{(A)_q}(z)(\partial^z_A)^q\delta(z-u)
\frac{\overleftarrow{\partial}}{\partial u^D}-\sigma(C,D)
(C\leftrightarrow D)=0
\]
and
\[
t_{BC}^{(A)_q}(z)(\partial^z_A)^q\delta (z-u)=
t_B^{(A)_{q-1}}(z)(\partial^z_A)^{q-1}\delta(z-u)
\frac{\overleftarrow{\partial}}{\partial u^C}.
\]
Hence
\[
\left[T_B(z,u)-\sum_{q=0}^Qt_B^{(A)_q}(z)(\partial^z_A)^q\delta(z-u)\right]
\frac{\overleftarrow{\partial}}{\partial u^C}=0
\]
and
\[
T_B(z,u)=\sum_{q=0}^Qt_B^{(A)_q}(z)(\partial^z_A)^q\delta(z-u)+\phi_B(z)=
\sum_{q=0}^Q\delta(z-u)(\overleftarrow{\partial}^u_A)^qt_B^{\prime(A)_q}(u)+
\phi_B(z).
\]
We therefore obtain $
\frac{\partial}{\partial z^C}T_B(z,u)-\sigma(B,C)(B\leftrightarrow C)=0\,,
$ or
$$
\frac{\partial}{\partial z^C}\left[\sum_{q=0}^Q\delta(z-u)
(\overleftarrow{\partial}^u_A)^qt_B^{\prime(A)_q}(u)+\phi_B(z)\right]-
\sigma(B,C)(C\leftrightarrow B)=0\,,
$$
which yields
\[
\frac{\partial}{\partial z^C}\sum_{q=0}\delta(z-u)
(\overleftarrow{\partial}^u_A)^qt_{B}^{\prime(A)_q}(u)-
\sigma(B,C)\frac{\partial}{\partial z^B}\sum_{q=0}\delta(z-u)
(\overleftarrow{\partial}^u_A)^qt_C^{\prime(A)_q}(u)=0
\]
and
\[
\frac{\partial}{\partial z^C}\phi_B(z)-
\sigma(B,C)\frac{\partial}{\partial z^B}\phi_C(z)=0\,.
\]
Thus
\[
T(z,u)=\sum_{q=0}t^{(A)_q}(z)(\partial^z_A)^q\delta(z-u)+\phi (z)+\psi(u).
\]
The relation
$T(z,u)=\frac{1}{2}(T(z,u)+(-1)^{n_-}T(u,z))$
gives
\[
T(z,u)=2\sum_{q=0}^Qt^{(A)_q}(z)(\partial^z_A)^q\delta(z-u)+\varphi(z)+
(-1)^{n_-}\varphi(u).
\]
Let
\[
m^\prime(z,u)=m(z,u)-(-1)^{n_-}\varphi(u).
\]
The distribution $m^\prime(z,u)$, which can be used instead of
$m(z,u)$, can be written as
\[
m^\prime(z,u)=m_-(z,u)+m_+(z,u)=m_2(z,u)+
\sum_{q=0}^Qt^{(A)_q}(z)(\partial^z_A)^q\delta(z-u),
\]
\[
m_\pm(z,u)=\frac{1}{2}[m^\prime(z,u)\pm(-1)^{n_-}m^\prime(u,z)],\quad
m_2(z,u)\equiv m_-(z,u)=-(-1)^{n_-}m_2(u,z).
\]
We now define the form
$M_{3|3}(f,g,h)$ by the relation
\[
M_{3|3}(f,g,h)=M_{3|2}(f,g,h)-d_2^{\rm tr}m_2(f,g,h).
\]
The form
$M_{3|3}(f,g,h)\equiv M_{3|{\rm loc}}(f,g,h)$
is absolutely local, i.e., it can take nonzero values only if
$
{\rm supp}(f)\bigcap{\rm supp}(g)\bigcap
{\rm supp}(h)\neq\varnothing.
$

Finally, we have
\[
M_3(f,g,h)=c\bar{f}\bar{g}\bar{h}+M_{3|{\rm loc}}(f,g,h)+
d_2^{\rm tr}m_2(f,g,h),
\]
\[
M_{3|{\rm loc}}(f,g,h)=\sum_{p,q,l=0}^NM^{(A)_p|(B)_q|(C)_l}([(\partial_A)^pf]
[(\partial_B)^qg][(\partial_C)^lh]),\;d_3^{\rm tr}M_{3|3\mathrm{loc}}=0.
\]

The solution of the cohomology equation for $M_{3|{\rm loc}}$ uses
the results of the next section and is described in Appendix~\ref{App7}. Up
to coboundaries, it has the following form
\begin{eqnarray}
M_{3|{\rm loc}}(f,g,h)=a\int dzf(z)\left[g(z)\left(
\frac{\overleftarrow{\partial}}{\partial z^A}\omega^{AB}
\frac{\partial}{\partial z^B}\right)^3h(z)\right]+ \nonumber   \\
b\int dz\{f(z),g(z)\}h(z), \label{4.11}
\end{eqnarray}
where
\[
a={\rm const},\quad \varepsilon(a)=n_-+\varepsilon(M_3),\quad
b={\rm const},\quad \varepsilon(b)=n_-+\varepsilon(M_3).
\]
One can easily check the antisymmetry property
\[
M_{3|{\rm loc}}(f,h,g)=-\sigma(g,h)M_{3|{\rm loc}}(f,g,h),\;
M_{3|{\rm loc}}(g,f,h)=-\sigma(f,g)M_{3|{\rm loc}}(f,g,h).
\]

\subsubsection{The case
$\mathbf {4+n_+-n_-=0\,.}$}\label{ii)}

In this case
\[
\hat{n}(z,u)\frac{\overleftarrow{\partial}}{\partial z^C}=0,\;n(z,u)=
(-1)^{\varepsilon_A}m^A(z,u)\frac{\overleftarrow{\partial}}{\partial z^A}\;
\Longrightarrow
\]
\[
\hat{n}(z,u)=-4\mu(u),\;\varepsilon(\mu)=\varepsilon(M_3)
\]
Let
\[
m^{A}(z,u)=\mu(u)z^A+m^{\prime A}(z,u),\quad\mu^\prime(u)=
(-1)^{\varepsilon_A}m^{\prime A}(z,u)\frac{\overleftarrow{\partial}}{\partial
z^A}=0\quad\Longrightarrow
\]
\[
\hat{m}^{\prime A}(z,u)\frac{\overleftarrow{\partial}}{\partial z^C}\omega^{CB}
-\sigma(A,B)\hat{m}^{\prime B}(z,u)
\frac{\overleftarrow{\partial}}{\partial z^C}\omega^{CA}=0,
\]
This equation can be solved analogously to equation (\ref{4.9}):
\begin{eqnarray}
M_{3|2}(f,g,h)=\gamma(f,g)\int dzz^A
\left(\frac{\partial f(z)}{\partial z^A}g(z)-
\sigma(f,g)\frac{\partial g(z)}{\partial z^A}f(z)\right)\mu(h)- \nonumber   \\
-\sigma(g,h)\gamma(f,h)\int dzz^A\left(\frac{\partial f(z)}{\partial z^A}h(z)-
\sigma(f,h)\frac{\partial h(z)}{\partial z^A}f(z)\right)\mu(g)+ \nonumber   \\
+\sigma(f,g)\sigma(f,h)\gamma(g,h)\int dzz^A\left(
\frac{\partial g(z)}{\partial z^A}h(z)-\sigma(g,h)
\frac{\partial h(z)}{\partial z^A}g(z)\right)\mu(f)+ \nonumber   \\
+m(\{f,g\},h)+\sigma(f,g)\sigma(f,h)m(\{g,h\},f)-
\sigma(g,h)m(\{f,g\},h)+ \nonumber   \\
+\sum_{p,q,l=0}^NM^{(A)_p|(B)_q|(C)_l}([(\partial_A)^pf]
[(\partial_B)^qg][(\partial_C)^lh]), \nonumber
\end{eqnarray}
\[
\gamma(f,g)=(-1)^{n_{-}+(n_{-}+\varepsilon(M_3))[\varepsilon(f)+
\varepsilon(g)]}.
\]

Substituting this expression in cohomology equation
(\ref{4.7})  and choosing such function $f$, $g$, $h$ and $s$ that
\[
\left[{\rm supp}(f)\bigcup{\rm supp}(g)\right]\bigcap\left[
{\rm supp}(h)\bigcup{\rm supp}(s)\right]=\varnothing,
\]
we obtain
\begin{eqnarray*}
\gamma(f,g)\int dzz^A\left(\frac{\partial f(z)}{\partial z^A}g(z)
-\sigma(f,g)\frac{\partial g(z)}{\partial z^A}f(z)\right)\mu(\{h,s\})+   \\
+\sigma(f,h)\sigma(f,s)\sigma(g,h)\sigma(g,s)\gamma(h,s)\int dzz^A\left(
\frac{\partial h(z)}{\partial z^A}s(z)-\right.   \\
\left.-\sigma(h,s)\frac{\partial s(z)}{\partial z^A}h(z)\right)\mu(\{f,g\})+   \\
+\hat{m}(\{f,g\},\{h,s\})+\sigma(f,h)\sigma(f,s)\sigma(g,h)\sigma(g,s)
\hat{m}(\{h,s\},\{f,g\})=0\,,
\end{eqnarray*}
which leads to
\begin{eqnarray*}
\int du\biggl(\biggl[2z^A\frac{\partial g(z)}{\partial z^A}-
4g(z))\mu(u)\{h(u),s(u)\}-    \\
-(-1)^{n_-}\sigma(g,h)\sigma(g,s)\sigma(M_3,h)\sigma(M_3,s)
(u^A\frac{\partial h(u)}{\partial u^A}s(u)-    \\
-\sigma(h,s)u^A\frac{\partial s(u)}{\partial u^A}h(u)\biggr]
\{\mu(z),g(z)\}+   \\
+\{\hat{m}(z,u),g(z)\}\{h(u),s(u)\}+(-1)^{n_-}\sigma(g,h)\sigma(g,s)
\{\hat{m}(u,z)\{h(u),s(u)\},g(z)\}\biggr)=0.
\end{eqnarray*}
Consider the terms proportional to
$g(z)$. We obtain
\[
\int du\mu(u)\{h(u),s(u)\}=0\;\Longrightarrow \;\mu(u)=\mu={\rm const}\;
\Longrightarrow
\]
\[
\int du[\{\hat{m}(z,u),g(z)\}\{h(u),s(u)\}+
(-1)^{n_-}\sigma(g,h)\sigma(g,s)\{\hat{m}(u,z)\{h(u),s(u)\},g(z)\}]=0.
\]
The last equation coincides with (\ref{4.10}), which is solved above.

\subsubsection{Independence and nontriviality}\label{InN}
Combining the results of the subsubsections \ref{i)} and \ref{ii)}, we finally obtain
\begin{eqnarray}
M_3(f,g,h)=c\bar{f}\bar{g}\bar{h}+a\int dzf(z)
\left[g(z)\left(\frac{\overleftarrow{\partial}}{\partial z^A}\omega^{AB}
\frac{\partial}{\partial z^B}\right)^3h(z)\right]+ \nonumber  \\
+\delta_{4+n_+,n_-}\mu U_3(f,g,h)+b\int dz\{f(z),g(z)\}h(z)+d_2^{\rm tr}M_2(f,g,h)\,,
 \label{4.12}
\end{eqnarray}
where
\begin{eqnarray*}
U_3(f,g,h)=(-1)^{n_{-}\varepsilon(f)}\bar{f}U_2(g,h)-
(-1)^{n_-\varepsilon(g)}\sigma(f,g)\bar{g}U_2(f,h)+   \\
+(-1)^{n_-\varepsilon(h)}\sigma(f,h)\sigma(g,h)\bar{h}U_2(f,g),   \\
U_2(f,g)=\int dz\left[z^A\frac{\partial f(z)}{\partial z^A}g(z)-
\sigma(f,g)z^A\frac{\partial g(z)}{\partial z^A}f(z)\right].
\end{eqnarray*}
Let us discuss the nontriviality and independence of these cocycles.
Consider two cases $n_+\ne n_-$ and $n_+ = n_-$ separately.

{\bf \ref{InN}.1 The case} $\mathbf {n_+\ne n_-}$

The first four terms in (\ref{4.12}) are independent nontrivial cocycles.
Indeed, let us suppose that
\begin{eqnarray}
\!\!\!\! c\bar{f}\bar{g}\bar{h}&+&a\int dzf(z)
\left[g(z)\left(\frac{\overleftarrow{\partial}}{\partial z^A}\omega^{AB}
\frac{\partial}{\partial z^B}\right)^3h(z)\right]+\nonumber \\
&+& b\int dz\{f(z),g(z)\}h(z)+
\delta_{4+n_+,n_-}\mu U_3(f,g,h)= \nonumber  \\
&=&d_2^{\rm tr}M_2(f,g,h)=M_2(\{f,g\},h)-\sigma(g,h)M_2(\{f,h\},g)-
M_2(f,\{g,h\}).  \label{4.13}
\end{eqnarray}
Let the supports of all three functions in (\ref{4.13}) do not
intersect. Then (\ref{4.13}) is reduced to
\[
c\bar{f}\bar{g}\bar{h}=0\quad\Longrightarrow c=0\quad\Longrightarrow
\]
\begin{eqnarray}
a\int dz f(z)\left[g(z)\left(
\frac{\overleftarrow{\partial}}{\partial z^A}\omega^{AB}
\frac{\partial}{\partial z^B}\right)^3h(z)\right]+b\int dz\{f(z),g(z)\}h(z)+
\nonumber   \\
+\delta_{4+n_+,n_-}\mu U_3(f,g,h)= \nonumber   \\
=d_2^{\rm tr}M_2(f,g,h)=M_2(\{f,g\},h)-\sigma(g,h)M_2(\{f,h\},g)-
M_2(f,\{g,h\}).  \label{4.14}
\end{eqnarray}
If the functions in (\ref{4.14}) are chosen in such a way that
\[
{\rm supp}(f)\bigcap{\rm supp}(h)={\rm supp}(g)\bigcap
{\rm supp}(h)=\varnothing,
\]
then (\ref{4.14}) gives
\[
\delta_{4+n_+,n_-}\mu(-1)^{n_-\varepsilon(h)}\sigma(f,h)\sigma(g,h)
\bar{h}U_2(f,g)=\hat{M}_2(\{f,g\},h),
\]
or, by varying w.r.t. $h(v)$ and taking into account that $n_-$ is
even:
\[
\mu\int dz\left[z^A\frac{\partial f(z)}{\partial z^A}g(z)-
\sigma(f,g)z^A\frac{\partial g(z)}{\partial z^A}f(z)\right]=
\int dz\hat{m}_2(z,v)\{f(z),g(z)\},\quad n_-=n_++4.
\]
Varying this equation w.r.t. $g(z)$, we have
\[
2\mu z^A\frac{\partial f(z)}{\partial z^A}-4\mu f(z)=
\{\hat{m}_2(z,v),f(z)\}.
\]
Considering the coefficient at
$f(z)$ in this equation we obtain
\[
\mu=0\quad\Longrightarrow
\]
\begin{eqnarray}
a\int dz f(z)\left[g(z)\left(
\frac{\overleftarrow{\partial}}{\partial z^A}\omega^{AB}
\frac{\partial}{\partial z^B}\right)^3h(z)\right]+
b\int dz\{f(z),g(z)\}h(z)= \nonumber  \\
=M_2(\{f,g\},h)-\sigma(g,h)M_2(\{f,h\},g)-M_2(f,\{g,h\}).  \label{4.15}
\end{eqnarray}
The equation (\ref{4.15}) is analyzed in the Appendix~\ref{App7}, where it is
shown that it has the solutions only if $a=b=0$. Thus the equation (\ref{4.13})
has the solution only if $c=\mu=a=b=0$.

{\bf \ref{InN}.2 The case} $\mathbf {n_+ = n_-\,.}$

In this case, the term $b\int dz\{f(z),g(z)\}h(z)$ is a trivial cocycle
(see Appendix~\ref{App7}). Thus, only the first three terms are nontrivial and
independent. The proof is completely analogous to the case $n_+\ne n_-$.

\section{Cohomologies in the adjoint representation} \label{adjo}

In this section, we prove Theorems~\ref{th2} and~\ref{th3}. We assume that the forms under
consideration take values in the space $\cal A$, where $\cal A$ is one of the
spaces $\mathbf D^{\prime n_-}_{n_+}$, $\mathbf E^{n_-}_{n_+}$, or $\mathbf
D^{n_-}_{n_+}$.

\subsection{$H^0_{\rm ad}$} \label{Had0}

The cohomology equation for the form $M_0(z)\in{\cal A}$ is given by \[
0=d_0^{\rm ad}M_0(z|f)=-\{M_0(z),f(z)\}\;\Longrightarrow \;M_{0}(z)=m_0={\rm
const}.\quad \]

If $M_0\in \mathbf D^{n_-}_{n_+}$, then we have $m_0=0$.

\subsection{$H^1_{\rm ad}$} \label{Had1}

{}For the form $M_1(z|f)=\int du\,m_1(z|u)f(u)\in{\cal A}$, the cohomology
equation has the form \begin{equation}
d_{1}^{\mathrm{ad}}M_{1}(z|f,g)=-\sigma(f,g)\{M_{1}(z|g),f(z)\}
+\{M_{1}(z|f),g(z)\}-M_{1}(z|\{f,g\})=0.  \label{5.1} \end{equation}

Suppose that 
$z\bigcap{\rm
supp}(f)=z\bigcap{\rm supp}(g)=\varnothing$ (here and subsequently, $z\cap
\ldots$ and $z\cup\ldots$ mean $\{x\}\cap\ldots $ and $\{x\}\cup\ldots$
respectively). Then we have $\hat{M}_1(z|\{f,g\})=0$ and,
therefore, $\hat{m}_1(z|u){\overleftarrow{\partial}}/{\partial u^A}=0$.
This equation is analogous to (\ref{4.2a}), and we have $$
m_1(z|u)=m_1(z)+\sum_{q=0}^Qt^{(B)_q}(z)(\partial^z_B)^q\delta (z-u),\quad
m_1(z)\in{\cal A}.  $$ Hence it follows that
$$M_1(z|f)=m_1(z)\bar{f}+\sum_{q=0}^Qt^{(B)_q}(z)(\partial^z_B)^qf(z)$$ (sign
factor, which does not depend on $f$, is omitted).

Let $z\bigcap{\rm supp}(g)={\rm supp}(f)\bigcap{\rm supp}(g)=\varnothing$.
Then  $\{f(z),m_1(z)\}\bar{g}=0$.  This means that $m_1(z)=m_1={\rm
const}$, $\,\varepsilon(m_1)=n_-+\varepsilon(M_1)$. If ${\cal A}=\mathbf
D^{n_-}_{n_+}$, then $m_1=0$. We thus have
\[M_1(z|f)=m_1\bar{f}+\sum_{q=0}^Qt^{(B)_q}(z)(\partial^z_B)^qf(z),\quad
\varepsilon(t^{(B)_q})=\varepsilon(M_1)+|\varepsilon_B|_{1,q}.  \]

To find the local part $$\sum_{q=0}^Qt^{(B)_q}(z)(\partial^z_B)^qf(z)$$ of
$M_1$, which in its turn satisfies (\ref{5.1}), we set $f=exp( z^A p_A)$ and
$g=exp(z^A k_A)$ in some neighbourhood of $z$.  Then the cohomology equation
for this local part takes the form \begin{equation} \lbrack
{}F(z|p)+F(z|k)-F(z|p+k)]\left[ p,k\right] +\{F(z|p),zk\}-\{F(z|k),zp\}=0,
\label{5.1a} \end{equation} where 
$$
{}F(z|p)=\sum_{q=0}^{Q}t^{(B)_{q}}(z)(p_{B})^{q},$$
$\left[p,k \right] =
(-1)^{\varepsilon _{A}}p_{A}\omega ^{AB}k_{B}=-\left[ k,p \right]$. Consider
the terms of the highest degree w.r.t $p$ and $k$ in this equation:
\begin{eqnarray}
[F_{Q}(z|p)+F_{Q}(z|k)-F_{Q}(z|p+k)]\left[ p,k\right] =0,
\label{5.1b}
\end{eqnarray}
where
$F_{Q}(z|p)=t^{(B)_{Q}}(z)(p_{B})^{Q}$. It
follows from Eq. (\ref{5.1b}) that $F_{Q}(z|p)=0,\;Q\geq 2$ or
$F(z|p)=t^{0}(z)+t^{B}(z)p_{B}$. Further, Eq. (\ref{5.1a}) gives
$t^{0}(z)\left[ p,k\right] +\{F(z|p),zk\}-\{F(z|k),zp\}=0$, or
$
t^{0}(z){\overleftarrow{\partial }}/{\partial z^{A}} =
0,$  
$t^{A}(z)\DD_z^B-\sigma (A,B)t^{B}(z) \DD_z^A = -t^{0}\omega
^{AB},$
which implies
$t^{0}(z)=t^{0}=\mathrm{const}$, $\varepsilon (t^{0})=\varepsilon (M_{1})$,
and $t^{A}(z)=-\frac{1}{2}t_{0}z^{A}-t(z)\DD_z^A$, $ \varepsilon
(t(z))=\varepsilon(M_{1})$.

Thus, the general solution of the cohomology equation (\ref{5.1}) is \[
M_1(z|f)=m_1\bar{f}+t^0 \eE_z f(z)-\{t(z),f(z)\}.  \] 
The first two terms in
this expression are independent nontrivial cocycles. Indeed, assume that
\begin{equation} m_1\bar{f}+t^0\eE_z f(z)=d_0^{\rm ad}q(z|f)=-\{q(z),f(z)\}.
\label{5.3} 
\end{equation} 
Let $z\bigcap{\rm supp}(f)=\varnothing$. Then
(\ref{5.3}) gives $m_1\bar{f}=0$. 
So $m_1=0$ and 
$$t^0\left(
1-\frac{1}{2}z^A\frac{\partial}{\partial z^A}\right)f(z)= -\{q(z),f(z)\}.$$
Considering the coefficient at $f(z)$ in this relation, we obtain $t^0f(z)=0$
or $t^0=0$, {\it i.e.}, the relation (\ref{5.3}) is satisfied only if
$m_1=t^0=0$.
Evidently, if $M_1(z|f) \in \mathbf D^{n_-}_{n_+}$ for any $f$, then $m_1=0$.

Let us discuss the term $-\{t(z),f(z)\}$. Is this expression a coboundary or
not? The answer depends on the functional class $\cal A$ in which the
considered multilinear forms take their values:

${\cal A}=\mathbf D^{\prime n_-}_{n_+}$. In this case $t(z)\in \mathbf
D^{\prime n_-}_{n_+}$ and the form $-\{t(z),f(z)\}$ is exact.

${\cal A}=\mathbf E^{n_-}_{n_+}$. In this case $t(z)\in \mathbf E^{\prime
n_-}_{n_+}$ and the form $-\{t(z),f(z)\}$ is exact.

${\cal A}=\mathbf D^{n_-}_{n_+}$. In this case the condition
$\{t(z),f(z)\}\in \mathbf D^{n_-}_{n_+}$ gives the restriction $ t(z)\in
\mathbf E^{n_-}_{n_+}$ only, and the form $-\{t(z),f(z)\}$ is exact if and
only if $t(z)\in \mathbf D^{n_-}_{n_+}$. So the forms $-\{t(z),f(z)\}$ are
independent nontrivial cocycles parameterized by the elements of quotient
space ${\raise2pt\hbox{$\mathbf E^{n_-}_{n_+}
$}}\big/{\raise-2pt\hbox{$\mathbf D^{n_-}_{n_+}$}}$.

\subsection{$H^2_{\rm ad}$} \label{Had2}

{}For the bilinear form \[ M_2(z|f,g)=\int dvdum_2(z|u,v)f(u)g(v)\in{\cal A},
\] the cohomology equation has the form \begin{eqnarray}
0&=&d_{2}^{\mathrm{ad}}M_{2}(z|f,g,h) =\nonumber \\
&=&\sigma(g,h)\{M_{2}(z|f,h),g(z)\}-\{M_{2}(z|f,g),h(z)\} -\sigma (f,gh)
 \{M_{2}(z|g,h),f(z)\}- \nonumber \\ &&-M_{2}(z|\{f,g\},h) +\sigma
(g,h)M_{2}(z|\{f,h\},g)+M_{2}(z|f,\{g,h\}).  \label{5.4} \end{eqnarray}

To solve Eq. (\ref{5.4}), we use the following sequence of propositions.

%{\bf Proposition 4.2.1}
\proposition\label{q4.2.1}
{\it
$M_2(z|f,g)=M_{2|1}(z|f,g)+M_{2|2}^1(z|f,g)+M_{2|2}^2(z|f,g)$, where
\begin{eqnarray} M_{2|1}(z|f,g)=c\bar{f}\bar{g},\;
\varepsilon(c)=\varepsilon(M_2),\quad d_{2}^{\rm ad}M_{2|1}(z|f,g,h)=0,
\label{5.4b}\\ M_{2|2}^{1}(z|f,g)=\sum_{q=0}^Q
m^{1(A)_q}(z|[(\partial^z_A)^qf(z)]g- \sigma(f,g)[(\partial^z_A)^qg(z)]f),
\label{5.4d}   \\
M_{2|2}^{2}(z|f,g)=\sum_{q=0}^Qm^{2(A)_q}(z|[(\partial_A)^qf]g-\sigma(f,g)
(\partial_A)^qg]f), \end{eqnarray} and the parities of the linear forms
$m^{1,2(A)_q}(z|\cdot)$ are given by \begin{eqnarray} \varepsilon
(m^{1,2(A)_q})=\varepsilon(M_2)+|\varepsilon_A|_{1,q}+n_-.  \end{eqnarray}
Due to antisymmetry of bilinear form the constant $c$ can be nonzero only if
$n_-$ is odd.}

To prove this proposition, it suffices to consider equation (\ref{5.4}) in
the following two domains:  
$$
z\bigcap\left[{\rm
supp}(f)\bigcup{\rm supp}(g)\bigcup{\rm supp}(h)\right]= {\rm
supp}(f)\bigcap\left[{\rm supp}(g)\bigcup{\rm supp}(h)\right]=\varnothing$$
and 
$$z\bigcap\left[{\rm supp}(f)\bigcup{\rm
supp}(g)\right]= {\rm supp}(f)\bigcap\left[{\rm supp}(g)\bigcup{\rm
supp}(h)\right]= {\rm supp}(g)\bigcap{\rm supp}(h)=\varnothing. \nonumber
$$

%{\bf Proposition 4.2.2}
\proposition\label{q4.2.2}
{\it  If $n_+\geq 4$ then \begin{eqnarray}
M_{2|2}^2(z|f,g)=\delta_{n_-,4+n_+}\mu(z)\int du\,\left( f(u)\eE_u
 g(u)-\sigma(f,g) g(u) \eE_u f(u)\right) +d_{1}^{{\rm ad}}\zeta(z|f,g)+...
\,\,,\nonumber \end{eqnarray} where the dots stand for the terms which can be
included into $M_{2|2}^1$.}

To prove this proposition, consider Eq. (\ref{5.4}) in the domain
$$z\bigcap\left[{\rm supp}(f)\bigcup{\rm supp}(g)\bigcup{\rm supp}(h)\right]=
\varnothing,$$ where $M_{2|2}^1(z|f,g)$ is identically zero. Using the results
of the subsection~\ref{Htr2}, we obtain $$ \hat{M}_{2|2}^2(z|f,g)=\int
du\,\hat{m}^{2A}(z|u)\left( \frac{\partial f(u)}{\partial u^A}g(u)-
\sigma(f,g)\frac{\partial g(u)}{\partial u^A}f(u)\right) $$ and the vector
$\hat{m}^{2A}(z|u)$ satisfies the equation $$
2\hat{m}^{2A}(z|u)\frac{\overleftarrow{\partial}}{\partial u^C}\omega^{CB}-
2\sigma(A,B)\hat{m}^{2B}(z|u)\frac{\overleftarrow{\partial}}{\partial u^C}
\omega^{CA}-\mu(z)\omega ^{AB}=0, $$ 
and $ \mu(z)=-m^{2A}(z|u)
\frac{\overleftarrow{\partial}}{\partial u^A}(-1)^{\varepsilon_A},\;
\varepsilon(\mu)=\varepsilon(M_2)+n_-, $ does not depend on $u$. It follows
from this equation that $ (4+n_{+}-n_{-})\mu(z) =0. $ Now we consider 2
cases:

i) $\mu(z)=0$. Since $n_+\geq 4$, the Poincare lemma implies that $
m^{2A}(z|u)=-\frac{1}{2}\zeta(z|u)\DD_z^A $ plus terms with support on the
plane $z=u$. We assign these terms to $M^1_{2|2}$.

ii) $4+n_{+}-n_{-}=0$. In this case, we have $
m^{2A}(z|u)=-\frac{1}{4}\mu(z)u^A-\frac{1}{2}\zeta(z|u) \DD_z^A $ plus terms
with support on the plane $z=u$. We assign these terms to $M^1_{2|2}$.

%{\bf Proposition 4.2.3}
\proposition\label{q4.2.3}
{\it If $n_+\geq 4$, then the kernels of the linear forms
$m^{1(A)_q}(z|\cdot )$ from Proposition~$\ref{q4.2.1}$ have the form
\begin{eqnarray} \hat{m}^{1(A)_0}(z|u)=a(u);\ \;\hat{m}^{1(A)_q}(z|u)=0
\mbox{ for }q\geq 2,\nonumber\\ m^{1A}(z|u)=-\frac{1}{2}a(u)z^{A}+m^1(z|u)
\frac{\overleftarrow{\partial}}{\partial z^C}\omega^{CA}+m^1_{\rm loc},\;
\varepsilon(m^1)=\varepsilon(M_2)+n_-,\nonumber
\end{eqnarray}
where $a(u)$ and $m^1(z|u)$ are some generalized functions and
$m^1_{\rm loc}$ is a generalized function with support in the plane
$z=u$. }

To prove this proposition, it suffices to consider
equation~(\ref{5.4}) in the domain $$\left[z\bigcup{\rm
supp}(f)\bigcup{\rm supp}(g)\right] \bigcap{\rm
supp}(h)=\varnothing$$ and note that this equation can be solved in
the same way as (\ref{5.1a}) if the variable $u$ is viewed as an
external parameter.

%{\bf Proposition 4.2.4}
\proposition\label{q4.2.4} {\it $\mu(z)$ from
Proposition~$\ref{q4.2.2}$ and $a(u)$ and $m^1(z,u)$ from
Proposition~$\ref{q4.2.3}$ have the form $a(u)=a={\rm const}$,
$\mu(z)=\mu={\rm const}$, and
\[
m^1(z|u)=\sum_{q=0}(\partial^z_C)^q\delta(z-u)t^{(C)_q}(u)+b(z)+c(u).
\]
}

Let $\left[z\bigcup{\rm supp}(h)\right]\bigcap\left[{\rm
supp}(f)\bigcup {\rm supp}(g)\right]=\varnothing$. Choosing
$h(u)={\rm const}$ in a neighborhood of $z$, we obtain $a(u)={\rm
const}$. Choosing $f(u)={\rm const}$ in a neighborhood of ${\rm
supp}(g)$, we find that $\mu(z)={\rm const}$. We therefore have
\[
\frac{\partial}{\partial z^A}\hat{m}^1(z|u)
\frac{\overleftarrow{\partial}}{\partial u^B}=0,
\]
and Proposition~\ref{q4.2.4} is proved.

Substituting the expression for $m^1(z|u)$ in the expression for
$M^1_{2|2}$, we find that the contribution of $c(u)$ is cancelled
and $b(z)$ gives a contribution only to pure differential.

The results of Propositions~\ref{q4.2.1} - \ref{q4.2.4} are
summarized in the following lemma.

%{\bf Lemma 4.2.1}
\lemma\label{4.2.1} {\it   If $n_+\geq 4$, then the general
solution of equation~$(\ref{5.4})$ has the form
\begin{eqnarray*}
M_2(z|f,g)= c\bar{f}\bar{g}+ a\int\!\!  du[\eE_z f(z)g(u)
-\sigma(f,g)\eE_z g(z)f(u)]+
\\ +\delta_{n_-,4+n_+}\mu\int du\,\left(f(u) \eE_u g(u)- \sigma(f,g)g(u)
\eE_u f(u)\right) +d_1^{\rm ad}\zeta^2(z|f,g)+M_{2|{\rm loc}}(z|f,g),   \\
\varepsilon(a)=\varepsilon(M_2)+n_-,\;\varepsilon(\mu)=\varepsilon(M_2)+n_-,
\end{eqnarray*}
where $c,\ a$, and $\mu$ are some constants and $c=0$ for even
$n_-$.

The local bilinear form $M_{2|{\rm loc}}(z|f,g)$ has the form
\begin{eqnarray}
\label{locform} M_{2|{\rm loc}}(z|f,g)=
\sum_{p,q=0}^N(-1)^{\varepsilon(f)|\varepsilon_B|_{1,q}}m^{(A)_p|(B)_q}(z)
[(\partial^z_A)^pf(z)](\partial^z_B)^qg(z),   \\ m^{(B)_q|(A)_p}=-
(-1)^{|\varepsilon_A|_{1,p}|\varepsilon_B|_{1,q}}m^{(A)_p|(B)_q},\
\ \
m^{(A)_0|(B)_0}=0,\\
\varepsilon(m^{(A)_p|(B)_q})=\varepsilon(M_2)+|\varepsilon_A|_{1,p}+
|\varepsilon_B|_{1,q},\nonumber
\end{eqnarray}
and satisfies the equation
\begin{equation}
\label{5.6} d_{2}^{{\rm ad}}M_{2|{\rm loc}}(z|f,g,h)=0.
\end{equation}
}

To solve equation~(\ref{5.6}), we use the method of the
work~\cite{Zh}. Let $x\in U$, where $U\subset {\mathbb R}^{n_+}$ is
some bounded domain. For the functions of the form
\[
f(z)=e^{z^Ak_A}f_0(z),\quad g(z)=e^{z^Al_A}g_0(z),\quad
h(z)=e^{z^Ar_A}h_0(z),\quad\varepsilon(k_A)=\varepsilon(l_A)=\varepsilon(r_A)
=\varepsilon_A,
\]
where $k$, $l$, and $r$ are new supervariables and
$f_0(z)=g_0(z)=h_0(z)=1$ for $x\in U$ ($z=(x,\xi)$),
equation~(\ref{5.6}) assumes the form
\begin{equation}
\label{general} \lbrack k,l]\Psi (z|k,l,r)+\{\psi
(z|k,l),zr\}+\mathrm{cycle}(k,l,r)=0,
\end{equation}
where
\begin{eqnarray} \Psi (z|k,l,r)=\psi (z|k+l,r)-\psi (z|k,r)-\psi (z|l,r),
\label{generala} \\ \psi (z|k,l)=
\sum_{p,q=0}^{N}m(z)^{(A)_{p}|(B)_{q}}(k_{A})^{p}(l_{B})^{q}= -\psi
(z|l,k)\,,  \label{generalb}
\end{eqnarray}
and the notation $\lbrack k,l]=(-1)^{\varepsilon _{A}}k_{A}\omega
^{AB}l_{B}$ introduced in the preceding subsection has been used.
By the antisymmetry property, we have $\psi(z|0,0)=0$. We note that
\begin{eqnarray}
\{f(z), z^Ap_A\}=f(z)\DD^A
p_A.
\end{eqnarray}
It is useful to write down the expression for the local form, which
is a coboundary (the differential of the 1-form with the kernel
$t^{(A)_{p}}(z)\partial_z^P\delta(z-u)$):
\begin{equation}\label{p-triv}
M_{2|\mathrm{triv}}(z|k,l)=\{t(z|k),zl\}-\{t(z|l),zk\}-T(z|k,l)[k,l],
\end{equation}
where
\[
T(z|k,l) =t(z|k+l)-t(z|k)-t(z|l)\quad \mbox{and}\quad t(z|k)
=\sum_{p=0}^{K}t^{(A)_{p}}(z)(k_{A})^{p}.
\]

We introduce the lowest degree filtrations $\oP_p$ and $\oP_{pq}$
of
polynomials.
\begin{eqnarray} && \mbox{\bf Definition. }\nonumber\\
&&\oP_{p}=\{f(k)\in \K [k_1,...,k_{n_++n_-}]:\nonumber\\ && \ \ \ \ \ \
\exists g\in \K [\alpha,k_1,...,k_{n_++n_-}]\ \ \ f(\alpha k)=\alpha^p
g(\alpha, k)\}  \,,       \\ &&\oP_{p,q}=\{f(k,l)\in \K
[k_1,...,k_{n_++n_-};l_1,...,l_{n_++n_-}]:  \nonumber\\ && \ \ \ \ \ \
\exists g   \in  \K [\alpha,\beta,k_1,...,k_{n_++n_-},l_1,...,l_{n_++n_-}]\ \
f(\alpha k,\beta l)=\alpha^p \beta^q g(\alpha, \beta, k, l)\}.
\end{eqnarray}

We obviously have $\oP_{p,q}\subset \oP_{r,s}$ for $p \geq r$ and
$q \geq s$. It is also clear that $fg\in \oP_{p+r,q+s}$ for $f\in
\oP_{p,q}$ and $g\in \oP_{r,s}$. Analogous relations are valid for
$\oP_p$.

To simplify equation~(\ref{general}), we represent the polynomial
$\psi(z|k,l)$ (\ref{generalb}) in the form
\begin{eqnarray}
\label{decomp} \psi(z|k,l)=t_0(z|k)-t_0(z|l)
+t^{AB}_{11}(z)k_A l_B + t^A_1(z|k) l_A - t^A_1(z|l) k_A + \varphi
(z|k,l),
\end{eqnarray}
where $t_0(z|k)\in \oP_1$, $t^A_1(z|k)\in \oP_2$, and
$\varphi(z|k,l)\in \oP_{2,2}$. We have the following propositions:

%{\bf Proposition 4.2.5}
\proposition\label{q4.2.5} {\it
$t_0(z|k)-t_0(z|l)=dt(z|k,l)+[k,l]t(z)$, where $t(z)\in {\cal A}$,
and the linear form $t(z|f)$ is given by $t(z|f)=-t(z)f(z)$. }

Setting $r=0$ in equation~(\ref{general}), we obtain
$[k,l](t_0(z|k+l)-t_0(z|k)-t_0(z|l))-\{t_0(z|k),
z^Al_A\}+\{t_0(z|l), z^Ak_A\}=0$. This equation coincides with
equation~(\ref{5.1a}) solved in the preceding section and its
solution is $t_0(z|k)=\alpha(1-(1/2) z^Ak_A)+\{t(z),\,z^Ak_A\}$. It
follows from the condition $t_0(z|0)=0$ that $\alpha=0$, which
implies the required statement.

%{\bf Proposition 4.2.6}
\proposition\label{q4.2.6} {\it If $n_+\geq 4$, then the term
$t_0(z|k)-t_0(z|l) +t^{AB}_{11}(z)k_A l_B$ in~$(\ref{decomp})$ is a
coboundary.}

By Proposition~\ref{q4.2.5}, we can write
$t_0(z|k)-t_0(z|l)+t^{AB}_{11}(z)k_A l_B= dt(z|k,l)+\tilde
t^{AB}_{11}(z)k_A l_B$. The terms in~(\ref{general}) linear in $k$,
$l$, and $r$ yield
\begin{eqnarray}
\sigma(C,A) \tilde t^{AB}_{11}(z)\DD^C+ \sigma(A,B) \tilde
t^{BC}_{11}(z)\DD^A+ \sigma(B,C) \tilde t^{CA}_{11}(z)\DD^B=0.
\end{eqnarray}
In view of the antisymmetry property $\tilde t^{AB}_{11}(z)=
-\sigma(A,B)\tilde t^{BA}_{11}(z)$ this equation implies for
$n_+\geq 4$ that $\sigma(A,B)\tilde t^{AB}_{11}(z)= p^A(z) \DD^B -
p^B(z)\DD^A$ with some vector-function $p^A(z)$. We thus obtain
$\tilde t^{AB}_{11}(z)k_A l_B=\{p(z|k),zl\}-\{p(z|l),zk\}$, where
$p(z|k)=p^A(z)k_A$. This coincides with~(\ref{p-triv}).

%{\bf Proposition 4.2.7}
\proposition\label{q4.2.7}
{\it Up to a coboundary, $t^A_1(z|k)l_A - t^A_1(z|l)
k_A\in \oP_{2,2}$}.

By Propositions~\ref{q4.2.5} and~\ref{q4.2.6}, we can assume
$t_0(z|k)-t_0(z|l) +t^{AB}_{11}(z)k_A l_B=0$. The terms
in~(\ref{general}) linear in $l$ and $r$ give the equation
$t^A_1(z|k)\DD^B - \sigma(A,B)t^B_1(z|k)\DD^A=0$ which implies that
$t^A_1(z|k)=t(z|k)\DD^A$ and $t^A_1(z|k) l_A - t^A_1(z|l) k_A=
\{t(z|k), zl\}-\{t(z|l),zk\}$. We therefore have $t^A_1(z|k) l_A -
t^A_1(z|l) k_A=\{t(z|k),zl\}-\{t(z|l),zk\}
-(t(z|k+l)-t(z|k)-t(z|l))[k,l]+(t(z|k+l)-t(z|k)-t(z|l))[k,l]= d
t(z|k,l)+\varphi_t (z|k,l)$, where $\varphi_t
(z|k,l)=(t(z|k+l)-t(z|k)-t(z|l))[k,l]\in \oP_{2,2}$. Thus, up to
coboundary, we have $\psi(z|k,l)\in \oP_{2,2}$.

{}From now on, we assume that $\psi(z|k,l)\in \oP_{2,2}$.

%{\bf Proposition 4.2.8}
\proposition\label{q4.2.8}
{\it The polynomial $\psi(z|k,l)$, $\psi(z|k,l)\in
\oP_{2,2}$, does not depend on $z$. }

Indeed, the linear in $r$ terms in equation~(\ref{general}) give
$\{\psi(z|k,l),\, zr\}=0$, which implies the statement of the
proposition.

Equation~(\ref{general}) is reduced to
\begin{equation}
[k,l]\Psi(k,l,r)+[l,r]\Psi(l,r,k)+[r,k]\Psi(r,k,l)=0,
\label{5.12}
\end{equation}
\begin{equation}
\Psi(k,l,r)=\psi(k+l,r)-\psi(k,r)-\psi(l,r),
\label{5.12a}
\end{equation}
where
\[
\psi(k,l)=\sum_{p,q=2}^Nm^{(A)_p|(B)_q}(k_A)^p(l_B)^q=-\psi(l,k).
\]

In the case when there are no odd variables ($n_-=0$) and $n_+\geq
4$, equation~(\ref{5.12}) was solved in \cite{Zh} and its general
solution has the form
\begin{equation}
\psi(k,l)=c([k,l])^3+[k,l]\bigl(\varphi(k+l)-\varphi(k)-\varphi(l)\bigr),
\label{5.13}
\end{equation}
where $c$ is an arbitrary constant and $\varphi(k)$ is an arbitrary
polynomial satisfying the condition\footnote{This form of the
general solution remains valid also for $n_+=2$, but this fact
requires a separate proof. Such a proof is given in~\cite{n=2}.}
$\varphi \in \oP_1$.

\subsubsection{The solution of equation~(\ref{5.12})}

%{\bf Theorem 4.3.1}
\theorem
 \label{th4.3.1}
{\it Let $n_+\geq 4$, $n_-\geq 0$, and $\psi(p,q)$ be a polynomial
with the property $\psi(p,q)=-\psi(q,p)\in\oP_{2,2}$. Let
\begin{equation}
\Psi(p,q,r)=\psi(p+q,r)-\psi(p,r)-\psi(q,r).
\label{4.3.1a}
\end{equation}
Then the general solution of the equation
\begin{equation}
[p,q]\Psi(p,q,r)+[q,r]\Psi(q,r,p)+[r,p]\Psi(r,p,q)=0
\label{eq}
\end{equation}
has the form
\begin{equation}
\psi(p,q)=c([p,q])^3+[p,q]\bigl(\varphi(p+q)-\varphi(p)-\varphi(q)\bigr),
\label{4.3.1b}
\end{equation}
where $[p,q]\stackrel {def}= (-1)^{\varepsilon _{A}}p_{A}\omega
^{AB}q_{B}$, $c$ is an arbitrary constant, and $\varphi(p)$ is an
arbitrary polynomial satisfying the condition $\varphi(p) \in
\oP_{1}$.}

The proof of this theorem can be obtained by induction on the
number of odd variables $n_-$ (see \cite{n=2}). Here we present a
proof similar to that given in \cite{Zh} for the purely even case.
This proof is based on the lemmas~\ref{A2.1} and~\ref{A2.2} of
Appendix~\ref{App3} and on the following two propositions.

%{\bf Proposition 4.3.1.1}
\proposition\label{q4.3.1.1} {\it If $n_+\geq 4$, then the solution
of equation~$(\ref{eq})$ has the form
\begin{equation} \label{prop}
\psi(p,q)=[p,q]g(p,q),
\end{equation}
where $g$ is a polynomial such that $g(p,q)=g(q,p)\in \oP_{1,1}$}.

{\it Proof}. By Lemma \ref{A2.2}, it follows from (\ref{eq}) that
\begin{eqnarray}\label{1-1}
\psi(p+q,r)-\psi(p,r)-\psi(q,r) = [r,p]Q
(p,q,r)+[q,r]P (p,q,r),
\end{eqnarray}
where $P$ and $Q$ are polynomials. We
apply the operation $\partial _{q_{A}}|_{q=0}$ to (\ref{1-1}).  We have
\begin{eqnarray}
\partial _{p_{A}}\psi(p,r) &=&r^{A}P (p,0,r)+[r,p]\partial
_{q_{A}}Q (p,q,r)|_{q=0}=  \nonumber \\
&=&\partial _{p_{A}}([p,r]P
(p,0,r))-[p,r]\partial _{p_{A}}P(p,0,r) +[r,p]\partial _{q_{A}}Q
(p,q,r)|_{p=0},
\end{eqnarray}
where $r^{A}=(-1)^A\omega
^{AB}r_{B}$. We hence obtain
\begin{eqnarray}  \label{9}
\partial_{p_A} (\psi(p,r)+[p,r]P(p,0,r))=
[p,r](-\partial_{p_A}P(p,0,r) +\partial_{q_A}Q(p,q,r)|_{q=0}).
\end{eqnarray}
Equation (\ref{9}) has the form
$\partial_{p_{A}}F(p,r)=[p,r]G^{A}(p,r)$, which implies that $N_p
{}F(p,r)=[p,r]p_A G^{A}(p,r)$, where $N_p=p_A\partial_{p_A}$ is the
Euler operator with respect to $p$. Obviously, the solution of this
equation has the form $F(p,r)=[p,r]H(p,r)+I(r)$, where $H$ and $I$
are some polynomials.  We therefore have
$\psi(p,r)=[p,r]g(p,r)+h(r)$ with some polynomials $g$ and $h$.
Because $\psi(0,r)=0$, we have $h(r)=0$ and hence
$\psi(p,r)=[p,r]g(q,r)$.

%{\bf Proposition 4.3.1.2}
\proposition\label{q4.3.1.2} {\it The polynomial $g(p,q)$ from
Proposition~$\ref{q4.3.1.1}$ has the form}
\begin{eqnarray}  \label{prop2}
g(p,q)= \varphi(p+q)-\varphi(p)-\varphi(q)+F(\left [p,q\right ]^2),
\end{eqnarray}
{\it where $\varphi$ and $F$ are some polynomials
such that $\varphi, F \in \oP_1$.}

{\it Proof}. Substituting expression~(\ref{prop}) in
equation~(\ref{eq}), we obtain
\begin{eqnarray}
\left [p,q\right ](\left [q,r\right ]-\left
[r,p\right ])W(p,q,r)= \left [r,p\right ](\left [p,q\right ]-\left
[q,r\right ])W(q,r,p),
\end{eqnarray}
where $W(p,q,r)=g(p+q,r)-g(q+r,p)+g(p,q)-g(q,r)$. Using Lemma
\ref{A2.1} twice, we find that
\begin{eqnarray}
W(p,q,r)=\left [r,p\right ]\left [r+p,q\right ]U(p,q,r),
\end{eqnarray}
where $U$ is a polynomial such that $U(p,q,r)=U(q,r,p)$. It follows
from the property $W(r,q,p)=-W(p,q,r)$ that $U(r,q,p)=U(p,q,r)$,
i.e., $U$ is a symmetric polynomial. Applying the operator
$\partial_{r_A}|_{r=0}$ to the relation
\begin{eqnarray} \label{formula}
g(p+q,r)-g(q+r,p)+g(p,q)-g(q,r)=\left [r,p\right
]\left [r+p,q\right ]U(p,q,r),
\end{eqnarray}
we obtain
$\varphi^A(p+q)-\partial_{q_A} g(p,q) - \varphi^A(q) = p^A \left
[p,q\right ] U(p,q,0)$, where
$\varphi^A(q)=\partial_{r_A}g(q,r)|_{r=0}$, and finally
\begin{eqnarray} \label{der}
\partial_{q_A} g(p,q)= \varphi^A(p+q)
- \varphi^A(q) - p^A \left [p,q\right ] U(p,q,0).
\end{eqnarray}
It follows from the condition $\partial_{q_B}\partial_{q_A} -
\sigma (A,B) \partial_{q_A}\partial_{q_B}=0$ that
\begin{eqnarray}
\!\!\!\!\!\!\!\!\!\!\!\!\!\!\!\!\!
\partial_{q_B}\varphi^A(p+q) - \partial_{q_B} \varphi^A(q) + \sigma(A,B) p^A
p^B U(p,q,0) - \sigma (A,B) p^A \left [p,q\right ] \partial_{q_B} U(p,q,0) =
\nonumber \\
=\left(\sigma (A,B) \partial_{q_A}\varphi^B(p+q) - \sigma
(A,B)\partial_{q_A} \varphi^B(q) + \right.\nonumber \\
\left. +p^B p^A U(p,q,0) - p^B \left [p,q\right ]
\partial_{q_A} U(p,q,0) \right),
\end{eqnarray}
and we have
\begin{eqnarray}
\label{eq2} \partial_{q_B}(\varphi^A(p+q) - \varphi^A(q))- \sigma (A,B)
\partial_{q_A}(\varphi^B(p+q) - \varphi^B(q)) =  \nonumber \\
=\left [p,q\right ](\sigma (A,B)  p^A \partial_{q_B} - p^B
\partial_{q_A}) U(p,q,0).
\end{eqnarray}
Let $q=0$. Then (\ref{eq2}) gives $\partial_{p_B}\varphi^A(p)
-\sigma (A,B)
\partial_{p_A}\varphi^B(p) =C_{BA} -\sigma (A,B)C_{AB}$, where $C_{AB}=
\partial_{p_A}\varphi^B(p)|_{p=0}=
\partial_{p_A}\partial_{q_B}g(p,q)|_{p=q=0}= \sigma(A,B) C_{BA}$ because
$g(p,q)=g(q,p)$. We thus obtain $\partial_{p_B}\varphi^A(p) -\sigma
(A,B)
\partial_{p_A}\varphi^B(p) =0$ and $\varphi^A(p)=
\partial_{p_A}\varphi(p)$, where $\varphi$ is a polynomial.
Equation~(\ref{eq2}) assumes the form
\begin{eqnarray}  \label{pseudo}
(p^A \partial_{q_B} -\sigma (A,B) p^B
\partial_{q_A}) U(p,q,0)=0
\end{eqnarray}
By Lemma \ref{A4}, it follows from
(\ref{pseudo}) that $\partial_{q_A}U(p,q,0)=p^A V(p,q)$, where $V(p,q)$ is a
polynomial, and in view of Lemma~\ref{A3} we have
\begin{eqnarray}
U(p,q,0)=H(\left [p,q\right ]),
\end{eqnarray}
where $H$ is a polynomial. Relation~(\ref{der}) now assumes the
form $\partial_{q_A} g(p,q)= \partial_{q_A}\varphi(p+q) -
\partial_{q_A}\varphi(q) - p^A \left [p,q\right ] H(\left [p,q\right ])$.
It hence follows that $ g(p,q)= \varphi(p+q) - \varphi(q)
-\varphi(p) - K(\left [p,q\right ]) + L(p)$ with some polynomials
$K$ and $L$. The condition $g(p,0)=0$ gives $L=0$, and the
condition $g(p,q)=g(q,p)$ gives $K=F(\left [p,q\right ]^2)$.

{\it Proof of Theorem~$\ref{th4.3.1}$.} We have established that
the solution of equation~(\ref{eq}) has the form
\begin{eqnarray}
\psi(p,q) = \left [p,q\right ]F(\left [p,q\right ]^2)+\psi_1(p,q),
\end{eqnarray}
where
\begin{eqnarray}
\psi_1(p,q)= \left [p,q\right
](\varphi(p+q) -\varphi(p) - \varphi(q)).
\end{eqnarray}
The polynomial $\psi_1$ satisfies equation~(\ref{eq}) for an
arbitrary polynomial $\varphi\in \oP_1$ and, therefore, $\left
[p,q\right ]F(\left [p,q\right ]^2)$ also satisfies this equation.
Obviously, the homogeneous parts of the polynomial $\psi=\left
[p,q\right ]F(\left [p,q\right ]^2)$ of all powers in $\left
[p,q\right ]$ should satisfy equation~(\ref{eq}). It is easy to
verify that $\psi=[p,q]^{2n+1}$ satisfies~(\ref{eq}) only if $n=1$.

Thus, the general solution $\psi(k,l)$ of equation~(\ref{5.12}) is
indeed of form~(\ref{5.13}) and the general solution of cohomology
equation~(\ref{5.4}) has the form
\begin{eqnarray}
M_2(z|f,g)&=&c_1f(z)\!\left(\frac{\overleftarrow{\partial}}{\partial z^A}
\omega^{AB}\frac{\partial}{\partial z^B}\right)^3\!g(z)+ c_2\bar{f}\bar{g}
+c_3[\eE_z f(z)\bar{g}- \sigma(f,g)\eE_z g(z)\bar{f}]+ \nonumber   \\
&+&\delta_{n_-,4+n_+}c_4\int du \left( f(u) \eE_u g(u)- \sigma(f,g) g(u)
\eE_u f(u)\right)+ d_1^{\rm ad}\zeta(z|f,g) \label{5.21c}.
\end{eqnarray}
It is easy to verify that $c_2=c_4=0$ if $M_2\in \mathbf
D^{n_-}_{n_+}$.

\subsubsection{Independence and nontriviality}

The first four terms in expression~(\ref{5.21c}) are independent
nontrivial cocycles. Indeed, suppose that
\begin{eqnarray}
c_1f(z)\!\left(\frac{\overleftarrow{\partial}}{\partial z^A}
\omega^{AB}\frac{\partial}{\partial
z^B}\right)^3\!\!g(z)+c_2\bar{f}\bar{g}+ c_3[\eE_z f(z)\bar{g}-
\sigma(f,g)\eE_z g(z)\bar{f}]\!+    \nonumber \\ +
    \delta_{n_-,4+n_+}c_4\!\!\int\!\!du\!\left( f(u)\eE_u g(u)-
    \sigma(f,g)g(u) \eE_u f(u)\!\right)\!= \nonumber \\ =d_1^{\rm
ad}\zeta(z|f,g)=-\sigma(f,q)\{\zeta(z|g),f(z)\}\! +
\{\zeta(z|f),g(z)\}\!-\zeta(z|\{f,g\}). \label{5.22}
\end{eqnarray}
Let the supports of the functions $f$ and $g$ satisfy the condition
$ z\bigcap{\rm supp}(f)=z\bigcap{\rm supp}(g)={\rm
supp}(f)\bigcap{\rm supp}(g) =\varnothing$. Then we have
$c_2\bar{f}\bar{g}=0\quad\Longrightarrow\quad c_2=0$. Further, if $
z\bigcap{\rm supp}(g)={\rm supp}(f)\bigcap{\rm
supp}(g)=\varnothing$, then we have $c_3\eE_z f(z)\bar{g}=
-\sigma(f,q)\{\zeta(z|g),f(z)\}$. This implies that $c_3=0$ and
$\hat{\zeta}(z|u)=a(u)$. Let $z\bigcap{\rm supp}(f)=z\bigcap{\rm
supp}(g)=\varnothing$. In this case, we obtain
\[
\delta_{n_-,4+n_+}c_4\int du \left( f(u) \eE_u g(u) - \sigma(f,g)
g(u) \eE_u f(u) \right)= -\zeta(z|\{f,g\}).
\]
Varying this equation with respect to $g(u)$, we find that
\[
\delta_{n_-,4+n_+}c_4\left(2u^A\frac{\partial f(u)}{\partial u^A}-
4f(u)\right)=\{\zeta(z|u),f(u)\},
\]
which implies that $c_4=0$ and
\[
\zeta(z|f)=a\bar f + \sum_{q=0}^Qt^{(D)_q}(z)(\partial^z_D)^qf(z).
\]

Equation~(\ref{5.22}) thus assumes the form
\begin{eqnarray}
-\sigma(f,g)\left\{\sum_{q=0}^Qt^{(D)_q}(z)(\partial^z_D)^qg(z),f(z)\right\}
+ \left\{
\sum_{q=0}^Qt^{(D)_q}(z)(\partial^z_D)^qf(z),g(z)\right\} - \nonumber   \\
-\sum_{q=0}^Qt^{(D)_q}(z)(\partial^z_D)^q\{f(z),g(z)\}=
c_1f(z)\left(\frac{\overleftarrow{\partial}}{\partial z^A}
\omega^{AB} \frac{\partial}{\partial z^B}\right)^3g(z).
\phantom{aa} \label{5.24}
\end{eqnarray}
We substitute the functions $f$ and $g$ of the form
\[
f(z)=e^{z^Ap_A}\varphi(x), \quad g(z)=e^{z^Ar_A}\phi(x)
\]
in equation~(\ref{5.24}), where $\varphi(x)=\phi(x)=1$ for $x$
belonging to some bounded domain $U\subset {\R}^{n_+}$
($z=(x,\xi)$). In this domain, equation~(\ref{5.24}) gives
\begin{equation}\label{zeta}
\{T(z|p),zr\}-\{T(z|r),zp\}+(T(z|p+r)-T(z|p)-T(z|r))[p,r]=-c_1[p,r]^3,
\end{equation}
where
\[
T(z|p)=\sum_{q=0}^Qt^{(D)_q}(z)(p_D)^q.
\]
Applying the operator $\overleftarrow\partial/\partial
r_A\biggr|_{r=0}$ to this equation yields
\[
T(z|p)\overleftarrow D^A=\tau^{AB}(z)p_B,
\]
where
\[
\tau^{AB}(z)=-T(z|0)(-1)^A \omega^{AB} +
T(z|r)\overleftarrow\partial/\partial
r_A\biggr|_{r=0}\overleftarrow D^B\quad \mbox{and}\quad
\overleftarrow D^A=\overleftarrow\partial/\partial z^B \omega^{BA}.
\]
As a consequence, we have $T(z|p)=\tau^A(z)p_A + T(p)$.
Substituting this expression in (\ref{zeta}), we see that the terms
belonging to $\oP_6$ give the equation
$(T(p+r)-T(p)-T(r))[p,r]=-c_1[p,r]^3$ or
${T(p+r)-T(p)-T(r)}=-c_1[p,r]^2$. Applying the operator
$\overleftarrow\partial/\partial r_A\biggr|_{r=0}$ to this equation
yields $\partial T(p)/\partial p_A=f^A$, where $f^A=\partial
T(r)/\partial r_A\biggr|_{r=0}$, which implies that $T(p)=F(0)+T^A
p_A$ and, therefore, $c_1=0$.

Thus, equation~(\ref{5.22}) has a solution only if
$c_1=c_2=c_3=c_4=0$.

\subsubsection{Exactness of the form $d_1^{\rm ad}\zeta(z|f,g)$}

We now discuss the term $d_1^{\rm ad}\zeta(z|f,g)$ in
expression~(\ref{5.21c}):
\begin{eqnarray*}
d_1^{\rm ad}\zeta(z|f,g)= \{\zeta(z|f),g(z)\}
-\sigma(f,g)\{\zeta(z|g),f(z)\} - \zeta(z|\{f,g\}).
\end{eqnarray*}
This form is a coboundary if $M_2(z|f,g)$ and $\zeta(z|g)$ belong
to the space ${\cal A}$.

1) ${\cal A}=\mathbf D^{\prime n_-}_{n_+}$. In this case, $\zeta(z|g)\in
\mathbf D^{\prime n_-}_{n_+}$ and the form $d_1^{\rm ad}\zeta(z|f,g)$ is
exact (trivial cocycle).

2) ${\cal A}=\mathbf E^{n_-}_{n_+}$. In this case, $\zeta(z|g)\in \mathbf
E^{n_-}_{n_+}$ and the form $d_1^{\rm ad}\zeta(z|f,g)$ is exact (trivial
cocycle).

3) ${\cal A}=\mathbf D^{n_-}_{n_+}$. In this case, the condition $d_1^{\rm
ad}\zeta(z|f,g)\in \mathbf D^{n_-}_{n_+}$ gives (see Appendix~\ref{App9})
\begin{equation}\label{repr}
\zeta(z|f)=\zeta_D(z|f)+\zeta(z)\bar{f},\quad
\zeta_D(z|f)\in \mathbf D^{n_-}_{n_+},\quad \zeta(z)\in \mathbf
E^{n_-}_{n_+}.
\end{equation}
The bilinear form $d_1^{\rm ad}\zeta(z|f,g)$ can therefore be
represented in the form $ d_1^{\rm
ad}\zeta(z|f,g)=M_{2|D}(z|f,g)+M_{2|C}(z|f,g)$, where
$M_{2|D}(z|f,g)=d_1^{\rm ad}\zeta_D(z|f,g)$ is an exact form and
$M_{2|C}(z|f,g)=\{f(z),\zeta(z)\}\bar{g}-
\sigma(f,g)\{\zeta(z),g(z)\}\bar{f}$ is a nontrivial cocycle
parameterized by the function classes in ${\raise2pt\hbox{$\mathbf
E^{n_-}_{n_+}$}} \big/{\raise-2pt\hbox{$\mathbf D^{n_-}_{n_+}$}}$.

\setcounter{equation}{0}
\def\theequation{A\arabic{appen}.\arabic{equation}}

\renewcommand{\theorem}{\par\refstepcounter{theorem}
{\bf Theorem A\arabic{appen}.\arabic{theorem}. }}
\renewcommand{\lemma}{\par\refstepcounter{lemma}
{\bf Lemma A\arabic{appen}.\arabic{lemma}. }}
\renewcommand{\proposition}{\par\refstepcounter{proposition}
{\bf Proposition A\arabic{appen}.\arabic{proposition}. }}
\makeatletter \@addtoreset{theorem}{appen}
\makeatletter \@addtoreset{lemma}{appen}
\makeatletter \@addtoreset{proposition}{appen}
\renewcommand\thetheorem{A\theappen.\arabic{theorem}}
\renewcommand\thelemma{A\theappen.\arabic{lemma}}
\renewcommand\theproposition{A\theappen.\arabic{proposition}}

\setcounter{equation}{0}
\appen \label{App2}

The space $\G^{n_{1-}}\otimes\ldots\otimes \G^{n_{k-}}$ is
naturally identified with $\G^{n_{1-}+\ldots+ n_{k_-}}$ by setting
\[
(g_1\otimes \ldots \otimes g_k)(\xi)=g_1(\xi_1)\ldots
g_k(\xi_k),\quad g_j\in \G^{n_{j-}},\,j=1,\ldots,k.
\]
Analogously,
for $\varphi_j\in \D(\R^{n_j})$ and $f_j\in \mathbf
D^{n_{j-}}_{n_{j+}}$, we set $$ (\varphi_1\otimes \ldots \otimes
\varphi_k)(x)=\varphi_1(x_1)\ldots \varphi_k(x_k), \quad
(f_1\otimes \ldots \otimes f_k)(z)=f_1(z_1)\ldots f_k(z_k).  $$ For
$g\in \G^{n_-}$ and $\varphi\in \D(\R^{n_+})$, the element
$\varphi\otimes g$ of the space $\mathbf D^{n_{-}}_{n_{+}}$ will be
denoted by $\varphi g$.

%\textbf{Lemma A1 }
\lemma\label{A1} {\bf (kernel theorem)}. \emph{ For every
separately continuous multilinear form $M$ on $\mathbf
D^{n_{1-}}_{n_{1+}}\times\ldots\times \mathbf D^{n_{k-}}_{n_{k+}}$,
there is a unique generalized function $m\in \mathbf D^{\prime
n_{1-}+\ldots+n_{k-}}_{n_{1+}+\ldots+ n_{k+}}$ such that
\begin{equation}\label{A2.1eq}
M(f_1,\ldots,f_k)=\int\di z_k\ldots\di z_1\,
m(z_1,\ldots,z_k)f_1(z_1)\ldots f_k(z_k)
\end{equation}
for any $f_1\in
\mathbf D^{n_{1-}}_{n_{1+}},\ldots, f_k\in \mathbf D^{n_{k-}}_{n_{k+}}$.}

\emph{Proof.} It suffices to consider the case $k=2$. The uniqueness of $m$
follows from the density of $\mathbf D^{n_{1-}}_{n_{1+}}\otimes \mathbf
D^{n_{2-}}_{n_{2+}}$ in $\mathbf D^{n_{1-}+n_{1-}}_{n_{1+}+n_{1+}}$ which is
ensured by the density of $\D^(\R^{n_{1+}})\otimes \D(\R^{n_{2+}})$ in
$\D(\R^{n_{1+}+n_{1+}})$. We now prove the existence of $m$. For every pair
$\varphi_{1}\in \D(\R^{n_{1+}})$, $\varphi_{2}\in \D(\R^{n_{2+}})$, we define
the continuous bilinear form $M_{\varphi_1,\,\varphi_2}$ on
$\G^{n_{1-}}\times \G^{n_{2-}}$ by the relation
\[
M_{\varphi_1,\,\varphi_2}(g_1,g_2)= M(\varphi_1 g_1,\varphi_2
g_2),\quad g_{1,2}\in \G^{n_{1,2-}}.
\]
This form uniquely determines a linear functional
$m_{\varphi_1,\,\varphi_2}$ on $\G^{n_{1-}+n_{2-}}$ such that
$m_{\varphi_1,\,\varphi_2}(g_1\otimes
g_2)=M_{\varphi_1,\,\varphi_2}(g_1,g_2)$. Clearly, the mapping
$(\varphi_1,\varphi_2)\to m_{\varphi_1,\,\varphi_2}(g)$ from
$\D(\R^{n_{1+}})\times \D(\R^{n_{2+}})$ to $\K$ is bilinear and
separately continuous for any $g\in \G^{n_{1-}+n_{2-}}$ and by the
standard kernel theorem for generalized functions (see,
e.g.,~\cite{Hoermander}, Theorem~5.2.1), there is a unique
generalized function $\mu_g\in \D'(\R^{n_{1+}+n_{2+}})$ such that
$\mu_{g}(\varphi_1\otimes \varphi_2)= m_{\varphi_1,\,\varphi_2}(g)$
for any $\varphi_{1,2}\in \D(\R^{n_{1,2+}})$. The density of
$\D(\R^{n_{1+}})\otimes \D(\R^{n_{2+}})$ in
$\D(\R^{n_{1+}+n_{1+}})$ implies that the mapping $g\to
\mu_g(\varphi)$ is linear in $g$ for any fixed $\varphi\in
\D(\R^{n_{1+}+n_{1+}})$. Hence, there is a unique $\mu\in
\D(\R^{n_{1+}+n_{1+}})\otimes \G^{n_{1-}+n_{2-}}$ such that
$\mu(\varphi g)=\mu_g(\varphi)$ for any $\varphi\in
\D(\R^{n_{1+}+n_{1+}})$ and $g\in \G^{n_{1-}+n_{2-}}$. For any
$f_1\in \mathbf D^{n_{1-}}_{n_{1+}}$ and $f_2\in \mathbf
D^{n_{2-}}_{n_{2+}}$, we have $\mu(f_1\otimes f_2)=M(f_1,f_2)$ and,
therefore, $m= L^{-1}\mu$ satisfies~(\ref{A2.1eq}) (here, $L$ is
the natural isomorphism between $\mathbf D^{\prime n_-}_{n_+}$ and
the space of continuous linear functionals on $\mathbf
D^{n_-}_{n_+}$).

\setcounter{equation}{0} \appen \label{App3}

In this appendix, we derive Lemmas~\ref{A2.1} and~\ref{A2.2} which
are used in the proof of Theorem~\ref{th4.3.1}. Perhaps the results
of this appendix follow from some general theory of rings of
polynomials in even and odd variables, but such results are unknown
to the authors. We formulate the lemmas with the minimal generality
necessary for our purposes.

Let $p$, $q$, and $r$ be three sets of supervariables with $n_+$
even and $n_-$ odd variables each. As above, the notation $[p,q]$
in this appendix means $[p,q]=(-1)^A\omega^{AB} p_A q_B$.

%{\bf Lemma A2.1}
\lemma\label{A2.1} {\it Let $n_+\geq 2$ and $n_-\geq 0$. If
$F(p,q,r)$ and $G(p,q,r)$ are polynomials in $p$, $q$, and $r$ then
the relation
\begin{eqnarray}
\label{LA3.1eq} [p,q]F+[q,r]G=0
\end{eqnarray}
is satisfied
if and only if there exists a polynomial $f(p,q,r)$ such that
\begin{eqnarray}\label{LA3.1sol}
F=[q,r]f(p,q,r),\ \ \ G=-[p,q]f(p,q,r).
\end{eqnarray}
}

{\it Proof}.  First we note that expression~(\ref{LA3.1sol})
satisfies equation~(\ref{LA3.1eq}) because the polynomials $[p,q]$
and $[q,r]$ are even. Secondly, it follows from~Lemma \ref{A2.3}
that Lemma~\ref{A2.1} is valid for $n_-=0$.

Suppose Lemma~\ref{A2.1} is valid for $n_-=N-1$. Let $P$, $Q$, and
$R$ be supervariables with $n_+$ even and $N$ odd components:
$P=(p,\mu_1)$, $Q=(q,\mu_2)$, $R=(r,\mu_3)$. Then we have
\begin{eqnarray*}
&&F(P,Q,R)= {}F_0(p,q,r)+ \mu_i F_i(p,q,r)+ \frac
1 2\mu_i\mu_j {}F_{ij}(p,q,r)+\mu_1\mu_2\mu_3 F_{123}(p,q,r),\\
&&G(P,Q,R)= G_0(p,q,r)+ \mu_i G_i(p,q,r)+ \frac 1
2\mu_i\mu_j G_{ij}(p,q,r)+\mu_1\mu_2\mu_3 G_{123}(p,q,r),\\
&&\left[P,Q\right]= [p,q]+\lambda_N\mu_1\mu_2,\ \ \
\left[Q,R\right]= [q,r]+\lambda_N\mu_2\mu_3
\end{eqnarray*}
(here, $\lambda_N=\pm 1$ is the parameter of the form
$\omega^{AB}$). By the induction hypothesis, equation
(\ref{LA3.1eq}) has the following solution:
\begin{eqnarray*}
F_0=[q,r]f_0,\ \ \ G_0=-[p,q]f_0\\ F_i=[q,r]f_i,\ \ \
G_i=-[p,q]f_i,\ \ \ i=1,2,3,\\ F_{12}=-F_{21}=[q,r]f_{12},\ \ \
G_{12}=-G_{21}=-[p,q]f_{12}-f_0, \\
F_{23}=-F_{32}=[q,r]f_{23}-f_0,\ \ \ G_{23}=-G_{32}=-[p,q]f_{23},
\\ F_{31}=-F_{13}=[q,r]f_{31},\ \ \ G_{31}=-G_{13}=-[p,q]f_{31},
\\ F_{123}=[q,r]f_{123}+f_1,\ \ \ G_{123}=-[p,q]f_{123}-f_3,
\end{eqnarray*}
where all functions $f_I$ ($I$ is an arbitrary index: $0$, $i$, and
so on) are polynomials in $p$, $q$, and $r$. Introducing the
polynomial $f(P,Q,R)=f_0+\mu_i f_i +1/2 \mu_i\mu_j
f_{ij}+\mu_1\mu_2\mu_3 f_{123}$, we can rewrite the obtained
solution in the form $F=[Q,R]f,\ G=-[P,Q]f$.

%{\bf Lemma A2.2}
\lemma\label{A2.2} {\it Let $n_+\geq 4$ and $n_-\geq 0$. If
$F(p,q,r)$, $G(p,q,r)$, and $H(p,q,r)$ are polynomials in $p$, $q$,
and $r$, then the relation
\begin{eqnarray}\label{LA3.2eq}
[p,q]F+[q,r]G+[r,p]G=0
\end{eqnarray}
is satisfied if and only if there exist polynomials
$f(p,q,r)$, $g(p,q,r)$, and $h(p,q,r)$ such that
\begin{eqnarray}\label{LA3.2sol}
F=[q,r]h-[r,p]g,\ \ \ G=[r,p]f-[p,q]h,\ \ \
H=[p,q]g-[q,r]f.
\end{eqnarray}
}

{\it Proof} is easily obtained by induction and is analogous to the
proof of the preceding lemma. The case $n_-=0$ is considered in
Lemma \ref{A2.4}.

%{\bf Lemma A2.3}
\lemma\label{A2.3} {\it Let $p$ be supervariable with $n\geq 2$
even components and no odd components. Let $F(p)$, $G(p)$, $u(p)$,
and $v(p)$ be polynomials in $p$ such that $u(p)$ and $v(p)$ do not
depend on $p_1$. Let $v|_{p_2=0}\neq 0$.  Then the relation
\begin{eqnarray}\label{LA3.3eq}
(p_1 p_2 + u) F+v G=0
\end{eqnarray}
is satisfied if and only if there is a polynomial $f(p)$ such
that
\begin{eqnarray}\label{LA3.3sol}
F=v f(p),\ \ \ G=-(p_1 p_2+u)f(p).
\end{eqnarray}
}

{\it Proof.} We perform the following change of variables: the
variable $x=p_1p_2+u$ is introduced instead of the variable $p_1$
and all other variables remain unchanged. Then we have
$p_1=(x-u)/p_2$. Let $\tilde F$ and $\tilde G$ denote the functions
$F$ and $G$ expressed in terms of the new variables. Relation
(\ref{LA3.3eq}) assumes the form $x\tilde F+v\tilde G=0$, where the
functions $\tilde F$ and $\tilde G$ are polynomials in $x$,
$p_2,...,p_{n}$, and $1/p_2$, i.e., $\tilde F, \tilde G \in
\K[x,p_2,...,p_{n},\,1/p_2]$.

We represent the function $\tilde G$ in the form $\tilde G=\tilde
G_{0}+x\tilde G_1$, where $\tilde G_1\in
\K[x,p_2,...,p_{n},\,1/p_2]$ and $\tilde G_0\stackrel {def} =
\tilde G|_{x=0} \in \K[p_2,...,p_{n},\,1/p_2]$. Then we have
$\tilde G_0=0$ and $\tilde F+v\tilde G_1=0$. Returning to the
initial variables, we obtain $F=-vG_1(p)$, where $G_1\in
\K[p,1/p_2]$. We expand $G_1$ in the powers of $1/p_2$:
\[
G_1=-f+\sum_{i=1}^N \frac{f_i}{p_2^i},
\]
where $f$ and $f_i$ are polynomials in $p$ such that $f_i$ do not
depend on $p_2$, $i=1,2,...$. Because $F$ is a polynomial and
$v|_{p_2=0}\neq 0$, we have $f_i=0$, $F=vf$, and $G_1=-f \in
\K[p]$. It hence obviously follows that $G=-(p_1p_2+u)f$.

%{\bf Lemma A2.4}
\lemma\label{A2.4}
{\it Let $p$, $q$, and $r$ be supervariables with $n\geq 4$
even components and with no odd components. If $F(p,q,r)$, $G(p,q,r)$, and
$H(p,q,r)$ are polynomials in $p$, $q$, and $r$ then the relation
\begin{eqnarray}\label{LA3.4eq}
[p,q]F+[q,r]G+[r,p]H=0
\end{eqnarray}
is satisfied if and only if there are polynomials $f(p,q,r)$,
$g(p,q,r)$, and $h(p,q,r)$ such that
\begin{eqnarray}\label{LA3.4sol}
F=[q,r]h-[r,p]g,\ \ \
G=[r,p]f-[p,q]h,\ \ \ H=[p,q]g-[q,r]f.
\end{eqnarray}
}

{\it Proof.} We introduce the new variables $x=[p,q]$, $z=[r,p]$,
and $u=q_1r_2 -q_2r_1$ instead of $p_1$, $p_2$, and $q_1$. All
other variables remain unchanged. Let
\[
y=[q,r],\ \ \ y_2=y-u=\sum_{i,j=3}^n \omega^{ij}q_i r_j,\ \ \
x_2=\sum_{i,j=3}^n \omega^{ij}p_i q_j,\ \ \ z_2=\sum_{i,j=3}^n
\omega^{ij}r_i p_j.
\]
Because $n\geq 4$, we have $y_2\neq 0$. The old variables are
expressed in terms of the new variables as follows:
\begin{eqnarray*}
\! p_1=-\frac {r_1r_2(x-x_2)+(q_2r_1+u)(z-z_2)}
{r_2u}, \  p_2=-\frac {r_2(x-x_2)+q_2(z-z_2)}{u},\  q_1=\frac
{q_2r_1+u}{r_2}.
\end{eqnarray*}
Relation~(\ref{LA3.4eq}) assumes the form
\begin{eqnarray*}
x\tilde
F + (u+y_2)\tilde G + z \tilde H=0,
\end{eqnarray*}
where $y_2\in\K[p_3,...,p_n,r_3,...,r_n]$, and $\tilde F,\tilde
G,\tilde H \in \K[x,z,u,p_3,...,p_n,\,q_2,...,q_n,\,r_1,...,r_n,
\,1/u,\\1/r_2]$. We represent the functions $\tilde G$ and $\tilde
H$ in the form
\begin{eqnarray}
\tilde{G}=\tilde{G}_0-x\tilde G_1,   \\
\tilde{H}=\tilde{H}_{0}-x\tilde H_1,
\end{eqnarray}
where $\tilde G_i, \tilde H_i \in
\K[x,z,u,p_3,...,p_n,\,q_2,...,q_n,\,r_1,...,r_n, \,1/u, \,1/r_2]$
and $\tilde{G}_0=\tilde{G}|_{x=0}$ and
$\tilde{H}_0=\tilde{H}|_{x=0}$ do not depend on $x$.

We obtain $\tilde F = -(u+y_2)\tilde G_1 -z\tilde H_1$. In the
initial variables, the obtained relation has the form
$F=(u+y_2)g-zh$, where $g,h\in \K[p,q,r,1/r_2,1/u]$. We expand $g$
and $h$ in the powers of $1/u$:
\[
g=g_0+\sum_{i=1}^{N}\frac {g_i} {u^i},\quad
h=h_0+\sum_{i=1}^{N}\frac {h_i} {u^i}.
\]
We obviously have $g_0, h_0\in\K[p,q,r,1/r_2]$ and $g_i, h_i \in
\K[p,q_2,...,q_n,r,1/r_2]$. Because $F$ is a polynomial, it follows
that $y_2 g_N-zh_N=0$ and $g_{i+1}+y_2 g_i-zh_i=0$, $i=1,...,N-1$.
The condition $n\geq 4$ implies that $y_2\neq 0$ and in view of
Lemma~\ref{A2.3} we conclude that $g_i=z\gamma_i$. We thus obtain
$F=yg_0+g_1-zh_0=yg_0-z(h_0-\gamma_1)$ or
\begin{equation}
F=y\mu+z\nu, \label{A3.7}
\end{equation}
where $\mu,\nu\in\K[p,q,r,1/r_2]$.

Let $M\geq1$ be the highest power in the expansion of $\mu$ and
$\nu$ in $1/r_2$, i.e.,
\[
\mu=\frac{\alpha_M}{r_2^M}+\mu_{M-1}^\prime,\quad
\nu=\frac{\beta_M}{r_2^M}+\nu_{M-1}^\prime,\quad \frac{\partial
\alpha_M}{\partial r_2}= \frac{\partial\beta_M}{\partial r_2}=0.
\]
Here, the functions $\alpha_M$ and $\beta_M$ are polynomials and
the functions $\mu_{M-1}^\prime$ and $\nu_{M-1}^\prime$ have the
degree $\leq M-1$ with respect to $1/r_2$. Relation (\ref{A3.7})
assumes the form
\begin{eqnarray*}
F=y\mu_{M-1}^\prime+z\nu_{M-1}^\prime+
\frac{q_1\alpha_M-p_1\beta_M}{r_2^{M-1}}+
\frac{y_0\alpha_M+z_0\beta _M}{r_2^M},   \\ y_0=-q_2r_1+y_2,\quad
z_0=r_1p_2+z_2.
\end{eqnarray*}
This relation implies the equation
$ y_0\alpha_M+z_0\beta_M=0. $ By Lemma~\ref{A2.3}, it has the
solution $ \alpha_M=z_0\varphi_M,\quad \beta_M=-y_0\varphi_M, $
where $\varphi_M$ is a polynomial. Using the relation $
q_1z_0+p_1y_0=q_1z+p_1y, $ we find that
\[
F=y\mu_{M-1}+z\nu_{M-1},\quad\mu_{M-1}=\mu_{M-1}^\prime+
\frac{p_1\varphi_M}{r_2^{M-1}},\quad\nu_{M-1}=\nu_{M-1}^\prime+
\frac{q_1\varphi_M}{r_2^{M-1}},
\]
where $\mu_{M-1}$ and $\nu_{M-1}$ are polynomials of degree $M-1$
with respect to $1/r_2$, and so on. We thus have verified that the
function $F$ can be represented in the form
\begin{equation}
{}F=yg-zh, \label{A3.8}
\end{equation}
where $g$ and $h$ are polynomials.

Analogously, we establish the validity of the following
representation for the functions $G$ and $H$: $G=zf_1+x(g_1-g),
\quad H=x(f_2-h)+y(g_2-h_1)$, where $f_1$, $f_2$, $g_1$, and $g_2$
are polynomials.

Substituting these expressions in equation~(\ref{LA3.4eq}) gives
$x(yg_1+zf_2)+yzg_2=0$. Applying Lemma~\ref{A2.3} several times, we
find that $g_1=z\varphi_1,\quad f_2=y\varphi_2,\quad
g_2=-x(\varphi_1+\varphi_2)$, which implies the statement of the
lemma with $f=f_1+x\varphi_1$.

\newpage
\setcounter{equation}{0} \appen \label{App4}

%{\bf Lemma A3}.
\lemma\label{A3}
{\it If $n_+\ge2$, then the general solution of the equation
\begin{equation}
\frac{\partial f(k,l)}{\partial k_A}=l^Ag(k,l),\ \ \
l^A=(-1)^{\varepsilon_A}\omega^{AB}l_B  \label{A4.1}
\end{equation}
for polynomials $f(k,l)$ and $g(k,l)$ has the form
\[
f(k,l)=H(x,l)=\sum_{q=0}^Qf_q(l)x^q,\quad g(k,l)= \frac{\partial
H(x,l)}{\partial x},
\]
where $f_q(l)$ are some polynomials in $l$ and $x=k_Al^A=[k,l]$.}

{\it Proof}. Multiplying equation~(\ref{A4.1}) by $k_A$, we obtain
$ N_kf(k,l)=xg(k,l)$, which implies that $f(k,l)=f_0(l)+xh_1(k,l)$,
where $h_1(k,l)$ is a polynomial. Substituting this representation
for the polynomial $f(k,l)$ in equation~(\ref{A4.1}) yields
$x\partial h_1(k,l)/\partial k_A=l^A[h_1(k,l)-g(k,l)]$. Considering
this equation for the case $A=1$, $\varepsilon_A=0$ and using
Lemma~\ref{A2.1}, we obtain $h_1(k,l)-g(k,l)=xg_1(k,l)$, where $
g_1(k,l)$ is a polynomial. It hence follows that $\partial
h_1(k,l)/\partial k_A=l^Ag_1(k,l)$ and, therefore,
$h_1(k,l)=f_1(l)+xh_2(k,l)$ and so on. Substituting the expression
$f(k,l)=H(x,l)$ in initial equation~(\ref{A4.1}), we obtain the
statement of the lemma for $g(k,l)$.

If $f(l,k)=\pm f(k,l)$, then all coefficient functions $f_q(l)$ do
not depend on $l$, and we can write $f(k,l)=H(x)$, where $H(x)$ is
a polynomial in $x$.

\setcounter{equation}{0} \appen \label{App6}

%{\bf Lemma A4}.
\lemma\label{A4} {\it If $n_+\neq 1$, then the general solution of
the equation
\begin{equation}
k_Aa_B(k)-\sigma(A,B)k_Ba_A(k)=0 \label{A6.1}
\end{equation}
for polynomials $a_A(k)$ has the form
\[
a_A(k)=k_Aa(k),
\]
where $a(k)$ is an arbitrary polynomial.}

{\it Proof.} Obviously, the expression $a_A(k)=k_Aa(k)$ is a
solution of equation~(\ref{A6.1}) for arbitrary functions $a(k)$.

We denote the even variables by $k_j$ ($j=1,...,n_+$) and the odd
variables by $k_\alpha$, $\alpha=1+n_+,...,n_-+n_+$. Applying the
operator $\sigma(A,B)\partial/\partial k_B$ to
equation~(\ref{A6.1}), we obtain $ [N_+ +n_+ +(n_- -
N_-)-(-1)^{\varepsilon_A}] a_A(k)=k_A \partial a_B(k)/\partial
k_B$, where $N_+=k_j\partial/\partial k_j$ and $N_-=
k_\alpha\partial/\partial k_\alpha$. Obviously, the operator $[N_+
+n_+ +(n_- - N_-)-(-1)^{\varepsilon_A}]$ is invertible in the space
of polynomials if $n_+\neq 1$ and has the only null eigenvector
$P(k)=k_2...k_{n_-+1}$ for $n_+=1$.

Thus, if $n_+\neq 1$ then we have $ a_A(k)=k_Aa(k)$, where
$a(k)=k_B [N_+ + n_+ +(n_- - N_-)] ^{-1} \partial a_B(k)/\partial
k_B$ is a polynomial.

If $n_+=1$, then there is the additional solution
\[
a_1(k)=k_2...k_{n_-+1},\ \ \ a_\alpha (k)=0, \ \
\alpha=2,...,n_-+1.
\]

\setcounter{equation}{0} \appen \label{App8}

In this appendix, we derive a useful representation for generalized
functions supported by a hyperplane.

%\textbf{Lemma A5.1.}
\lemma\label{A5.1} \emph{Let $u\in\D'(\R^n)$ be a generalized
function of order $\leq N$ with compact support, $f\in
C^\infty(\R^n)$, and $\partial^\lambda f(x)=0$ for $|\lambda|\leq
N$ and $x\in\supp u$. Then we have $u(f)=0$. }

The proof of Lemma~\ref{A5.1} can be found in \cite{Hoermander}.

%\textbf{Lemma A5.2.}
\lemma\label{A5.2} \emph{For every generalized function $u\in
\D'(\R^{n_{1}+n_{2}})$ whose support is contained in the hyperplane
$x_2=0$, there is a unique locally finite decomposition
\begin{equation}\label{1}
u=\sum_{\lambda}
u_{\lambda}(x_1)\partial^\lambda\delta(x_2),
\end{equation}
where
$u_\lambda\in\D'(\R^{n_1})$ and the summation is performed over all
multi-indices $\lambda=(\lambda_1,\ldots,\lambda_{n_2})$.}

\emph{Proof.} Let $\varphi\in \D(\R^{n_2})$ and $\varphi=1$ in a
neighborhood of the origin. For every multi-index $\lambda$, we
define the linear operator $L_\lambda: \D(\R^{n_1})\to
\D(\R^{n_1+n_2})$ by the relation $[L_\lambda
f](x_1,x_2)=(-1)^{|\lambda|}f(x_1)\varphi(x_2)x_2^\lambda/\lambda!$,
$f\in\D(\R^{n_1})$. If a family of generalized functions
$u_\lambda$ satisfies (\ref{1}), then we have
\begin{equation}\label{2}
u_\lambda(f)=u(L_\lambda f)
\end{equation}
for any $f\in\D(\R^{n_1})$ and, therefore, representation (\ref{1})
is unique. We now show that the generalized functions $u_\lambda$
defined by (\ref{2}) form a locally finite family satisfying
(\ref{1}). Let $K$ be a compact set in $\R^{n_1}$ and $\chi$ be a
function in $\D(\R^{n_1})$ equal to unity in a neighborhood of $K$.
Let $u^\chi$ be the generalized function with compact support
defined by the relation $u^\chi(x_1,x_2)=\chi(x_1)u(x_1,x_2)$ and
let $N$ denote the order of $u^\chi$. If $|\lambda|>N$ and $f\in
\D(\R^{n_1})$, then all derivatives of $L_\lambda f$ of order $\leq
N$ vanish on the support of $u^\chi$ and by Lemma~\ref{A5.1}, we
have $u^\chi(L_\lambda f)=0$. For $\supp f\subset K$, we hence
obtain
\[
u_\lambda(f)=u(L_\lambda f)=u^\chi(L_\lambda f)=0.
\]
The family $u_\lambda$ is therefore locally finite. It remains to
show that the relation
\begin{equation}\label{3}
u(g)=\sum_{|\lambda|\leq N}(-1)^{|\lambda|}u(L_\lambda g_\lambda),
\end{equation}
where $g_\lambda(x_1)=\partial^\lambda g(x_1,x_2)|_{x_2=0}$, holds
for any $g\in \D(\R^{n_1+n_2})$ such that $\supp g\subset
K\times\R^{n_2}$. We observe that the function
$G=\sum_{|\lambda|\leq N}(-1)^{|\lambda|} \tilde g_\lambda$ is, up
to the factor $\varphi(x_2)$, the Taylor expansion of the function
$g$ in $x_2$-variable. Therefore, all derivatives of the function
$g-G$ of order $\leq N$ vanish on the support of $u_\chi$ and we
have $u_\chi(g-G)=0$ by Lemma~\ref{A5.1}. Hence the relation
(\ref{3}) follows because $u(g-G)=u_\chi(g-G)$ by the above
assumption about the support of $g$.

\setcounter{equation}{0} \appen \label{App9}

In this Appendix, we prove the existence of
representation~(\ref{repr}) for every $\zeta(z|f)\in \mathbf
E^{n_-}_{n_+}$ such that $\di^{\mathrm{ad}}_1 \zeta(z|f,g)\in
\mathbf D^{n_-}_{n_+}$ for any fixed $f,g\in \mathbf
D^{n_-}_{n_+}$. Obviously, it suffices to find $\zeta_D(z|f)\in
\mathbf D^{n_-}_{n_+}$ such that
\[
\frac{\partial \zeta(z|u)}{\partial u^A}= \frac{\partial
\zeta_D(z|u)}{\partial u^A},\quad A=1,\ldots, n_++n_-.
\]
It immediately follows from expression~(\ref{5.1}) for the
differential $\di^{\mathrm{ad}}_1$ that $\zeta(z|\{f,g\})\in
\mathbf D^{n_-}_{n_+}$ for any fixed $f,g\in \mathbf
D^{n_-}_{n_+}$. This implies that $\zeta(z|\partial f/\partial
u^A)\in \mathbf D^{n_-}_{n_+}$ for any $f\in \mathbf
D^{n_-}_{n_+}$, and the existence of $\zeta_D(z|f)$ with the
required properties is ensured by the following lemma (for
simplicity, we formulate this lemma for the case $n_-=0$).

%\textbf{Lemma A6}
\lemma\label{A6} \emph{Let generalized functions $T_i\in
\D'(\R^{n_1+n_2})$, $i=1,\ldots,n_1$, satisfy the conditions
\begin{equation}\label{4}
\frac{\partial T_i}{\partial
x^j}(x,y)=\frac{\partial T_j}{\partial x^i}(x,y),\quad
i,j=1,\ldots,n_1,
\end{equation}
and let the function $\varphi\in\D(\R^{n_1})$ be such that
\[
\int_{\R^{n_1}} \varphi(x)\,\di x=1.
\]
{}For $\psi\in \D(\R^{n_1+n_2})$, set
\[
T(\psi) = \sum_{i=1}^{n_1} T_i(\psi^i),
\]
where
\begin{equation}\label{5}
\psi^i(x,y)= \int_0^1\di
t\int_{\R^{n_1}}\di x' x^{\prime i}
\varphi(x-tx')\psi(x+(1-t)x',y).
\end{equation}
Then the functional $T$ belongs to $\D'(\R^{n_1+n_2})$ and
satisfies the conditions $T_i=\partial T/\partial x^i$ for every
$i=1,\ldots,n_1$. }

\emph{Proof.} Obviously, the integral in the right-hand side of
(\ref{5}) is convergent for any $x$ and $y$, and $\psi^i$ are
$C^\infty$-functions on $\R^{n_1+n_2}$. We show that
$\psi^i\in\D(\R^{n_1+n_2})$. Let $R>0$ be such that
$\supp\varphi\subset\{\,x\in \R^{n_1}\,|\,|x|\leq R\,\}$ and $\supp
\psi\subset K_R= \{\,(x,y)\in \R^{n_1+n_2}\,|\,\max(|x|,|y|)\leq
R\,\}$. It follows from~(\ref{5}) that $\psi^i$ can be nonzero only
if
\[
|x-tx'|\leq R,\quad |x+(1-t)x'|\leq R, \quad |y| \leq R.
\]
This implies that $|x'|\leq 2R$ and, therefore, $|x|\leq 3R$. We
thus have $\supp \psi^i\subset K_{3R}$ and
$\psi^i\in\D(\R^{n_1+n_2})$. Let $\D(K)$ denote the space of smooth
functions whose supports are contained in a compact set $K$. As
shown above, $\psi^i\in \D(K_{3R})$ for $\psi\in \D(K_R)$. We now
check that the mappings $L^i: \D(\R^{n_1+n_2})\to \D(\R^{n_1+n_2})$
taking $\psi$ to $\psi^i$ are continuous and, as a consequence,
$T\in \D'(\R^{n_1+n_2})$. It suffices to show that $L^i$ are
continuous as mappings from $\D(K_{R})$ to $\D(K_{3R})$ for every
$R>0$, i.e., that for any $\psi\in \D(K_R)$ and every
$k=0,1,\ldots$, the inequality $\|\psi^i\|_k\leq C\|\psi|_{k'}$
holds, where
\[
\|\psi\|_k = \max_{|\lambda|\leq k} \sup_{x,y}
|\partial^\lambda\psi(x,y)|
\]
and $C$ and $k'$ depend only on $k$ and $R$ (but not on $\psi$).
For $|\lambda|\leq k$, we have
\[
|\partial^\lambda\psi^i(x,y)|\leq \int_0^1\,\di t\int_{|x'|\leq
2R}\,\di x' |x^{\prime
i}\partial^\lambda[\varphi(x-tx')\psi(x+(1-t)x',y)]|\leq 2^k
\int_{|x'|\leq 2R}\di x'\, \|\varphi\|_k\|\psi\|_k.
\]
Hence $\|\psi^i\|_k\leq C\|\psi\|_k$, where
\[
C=2^k \int_{|x'|\leq 2R}\di x'\, \|\varphi\|_k.
\]

We now verify that $T_i=\partial T/\partial x^i$. Performing the
change of variables $x'\to (x-s)/t$ in the integral representing
$L^i\partial\psi/\partial x^j$, we find that
\[
\left[L^i\frac{\partial\psi}{\partial x^j}\right]\!\!(x,y) = \!
\int_0^1\!\!\di t\!\int_{\R^{n_1}}\! \di s
\frac{x^i-s^i}{t^n}\varphi(s)\frac{\partial}{\partial x^j}\!
\left[\psi\left(\frac{x+(t-1)s}{t},\,y\right)\!\right]\!=
\!\frac{\partial}{\partial x^j}\Phi^i(x,y) - \delta^i_j\Phi(x,y), \nonumber \\
\]
where
\begin{eqnarray} \Phi^i(x,y)& =& \int_0^1\di
t\int_{\R^{n_1}} \di s \frac{x^i-s^i}{t^n} \varphi(s)
\psi\left(\frac{x+(t-1)s}{t},\,y\right)= \nonumber \\ \ \ \ \ \ \ \
&=& \int_0^1 t\,\di t\int_{\R^{n_1}}\di x' x^{\prime i}
\varphi(x-tx')\psi(x+(1-t)x',y), \label{6}\\
\Phi(x,y) & = &
\int_0^1\di t\int_{\R^{n_1}} \frac{\di s}{t^n} \varphi(s)
\psi\left(\frac{x+(t-1)s}{t},\,y\right)= \nonumber \\ \ \ \ \ & =&
\int_0^1 \di t\int_{\R^{n_1}}\di x'
\varphi(x-tx')\psi(x+(1-t)x',y). \label{7}
\end{eqnarray}
As in the case of $L^i\psi$ above, we make sure that
$\Phi,\,\Phi^i\in \D(\R^{n_1+n_2})$. We therefore have
\begin{eqnarray*}
\left[\frac{\partial T}{\partial
x^j}\right](\psi)=- T\left(\frac{\partial \psi}{\partial
x^j}\right)= -\sum_{i=1}^{n_1} T_i\left(L^i\frac{\partial
\psi}{\partial x^j}\right)= -\sum_{i=1}^{n_1}
T_i\left(\frac{\partial
\Phi^i}{\partial x^j}\right)+ T_j(\Phi)=                    \\
=T_j\left(-\sum_{i=1}^{n_1} \frac{\partial \Phi^i}{\partial x^i}+
\Phi\right),
\end{eqnarray*}
where (\ref{4}) has been used for deriving the last equality. But
it follows from (\ref{6}) and (\ref{7}) that
\begin{eqnarray*}
-\sum_{i=1}^{n_1} \frac{\partial \Phi^i}{\partial
x^i}(x,y)=\int_0^1\di t\, t\frac{\di}{\di t} \int_{\R^{n_1}} \di x'
\varphi(x-tx')\psi(x+(1-t)x',y)=\\
= \psi(x,y)\int \di x' \varphi(x-x')
-\Phi(x,y)=\psi(x,y)-\Phi(x,y).
\end{eqnarray*}
We hence have $\left[\partial T/\partial
x^j\right](\psi)=T_j(\psi)$.

\setcounter{equation}{0}
\appen
\label{App7}

We discuss here the solution of the cohomology equation (\ref{4.7}) for
$M_{3|{\rm loc}}$. In this Appendix we write the upper indices
``(tr)'' or ``(ad)'' at corresponding forms.

Let us establish a correspondence between the forms $M^{({\rm tr})}_{p+1}$
and $M^{({\rm ad})}_p$. Namely, let
\[
M^{({\rm tr})}_{p+1}(f,g,h,\ldots)=
\int\ldots dvdudzm_{p+1}^{({\rm tr})}(z,u,v,\ldots)f(z)g(u)h(v)\cdots,
\]
and
\[
M^{({\rm ad},a)}_p(z|g,h)=
\int\ldots dvdum_p^{({\rm ad},a)}(z|u,v,)g(u)h(v)\cdots\,.
\]
Then the correspondence is defined by the formula
\begin{equation}
m_p^{({\rm ad},a)}(z|u,v,\ldots)=m_{p+1}^{({\rm tr})}(z,u,v,\ldots),\quad
\varepsilon(M^{({\rm ad},a)}_p)=\varepsilon(M^{({\rm tr})}_{p+1})+n_- \,,
\label{A7.1}
\end{equation}
where upper index ``a'' means that the kernel
$m_p^{({\rm ad},a)}(z|u,v,\ldots)$ is completely antisymmetric under
transpositions of all coordinates including $z$. Such forms
$M^{({\rm ad},a)}_p$ are called here completely antisymmetric. Conversely,
having completely antisymmetric form $M^{({\rm ad},a)}_p$, we can construct
the form $M^{({\rm tr})}_{p+1}$.  with the help of the same correspondence
(\ref{A7.1}) between the kernels.  In such a way, there exists a one-to-one
correspondence between the forms $M^{({\rm tr})}_{p+1}$ and completely
antisymmetric forms $M^{({\rm ad},a)}_p$. The following relation
\[
(-1)^{n_-p}\sigma(f,M^{({\rm ad},a)}_p)\int dzf(z)
M^{({\rm ad},a)}_p(z|f_1,\ldots)=M^{({\rm tr})}_{p+1}(f,f_1,\ldots)
\]
takes place.

The form $M^{({\rm tr})}_p$ is called local $M^{({\rm tr})}_{p|{\rm loc}}$,
if it equals zero in each case when supports of some two arguments do not
intersect. Analogously the form $M^{({\rm ad})}_p(z|\ldots)$ is called local
$M^{({\rm ad})}_{p|{\rm loc}}(z|\ldots)$, if it equals zero in each case
when supports of some two function-arguments do not intersect or when the
support of some function do not intersect with the point $z$.

Evidently, the relation (\ref{A7.1}) defines the one to one correspondence
between local forms $M^{({\rm tr})}_{p+1|{\rm loc}}$ and
$M^{({\rm ad},a)}_{p|{\rm loc}}$.

Let us establish the connection
between the actions of differentials on $M^{({\rm ad},a)}_p$ and
$M^{({\rm tr})}_{p+1}$.

The expression for $d^{\rm ad}_pM^{({\rm ad},a)}_p$ is defined by formula
(\ref{3.2c}). Apply the operator
$(-1)^{n_-p}\sigma(f,M^{({\rm ad},a)}_p)\int dzf(z)$ to this expression.
In view of (\ref{3.0aa}) the first term of the first sum in~(\ref{3.2c}) is
rewritten as
\begin{eqnarray*}
(-1)^{n_-p}\sigma(f,M^{({\rm ad},a)}_p)\!\int dzf(z)
\sigma(f_1,M^{({\rm ad},a)}_p)\{f_1(z),\!
\int\!\!\ldots dz_1m^{({\rm ad},a)}_p(z|z_1,\ldots)f_2(z_1)\cdots\}= \\
=\!(-1)^{n_-p}\sigma(f,M^{({\rm ad},a)}_p)\sigma(f_1,M^{({\rm ad},a)}_p)\!\!\!
\int\!\! dz\{f(z),f_1(z)\}\!\!\!\int\!\!\ldots dz_1
m^{({\rm ad},a)}_p(z|z_1,\ldots) f_2(z_1)\cdots\!= \\
=\int\ldots dz_1dzm^{({\rm ad},a)}_p(z|z_1,\ldots)\{f(z),f_1(z)\}
f_2(z_1)\cdots,
\end{eqnarray*}
which exactly coincides with the first term ($i=1$, $j=2$) in the expression
(\ref{3.2b}) for $d^{\rm tr}_{p+1}M^{({\rm tr})}_{p+1}$.
For the other terms, we similarly obtain
\[
(-1)^{n_-p}\sigma(f,M^{({\rm ad},a)}_p)\int dzf(z)d^{\rm ad}_p
M^{({\rm ad},a)}_p(z|f_1,\ldots)=
d^{\rm tr}_{p+1}M^{({\rm tr})}_{p+1}(f,f_1,\ldots).
\]
It hence follows that the relation~(\ref{A7.1})
determines
a one-to-one correspondence between the closed forms (cocycles)
$M^{({\rm ad},a)}_p$ and $M^{({\rm tr})}_{p+1}$:
\[
d^{\rm ad}_pM^{({\rm ad},a)}_p=0\quad\Longleftrightarrow\quad
d^{\rm tr}_{p+1}M^{({\rm tr})}_{p+1}=0.
\]

If the form $M^{({\rm tr})}_{p+1}$ is exact (coboundary),
$M^{({\rm tr})}_{p+1}=d^{\rm tr}_pM^{({\rm tr})}_p$, then the corresponding
form $M^{({\rm ad},a)}_p$ is also exact,
$M^{({\rm ad},a)}_p=d^{\rm ad}_{p-1}M^{({\rm ad},a)}_{p-1}$, where
$M^{({\rm ad},a)}_{p-1}$ is constructed from $M^{({\rm tr})}_p$ by
(\ref{A7.1}).

The converse statement is not valid.

We now solve equation (\ref{4.7}) for the form
$M^{({\rm tr})}_{3|{\rm loc}}$
using the general form (\ref{4.3.1b})
of the solution of the cohomology equation (\ref{5.6})
for the form $M^{({\rm ad})}_{2|{\rm loc}}$. From the above it follows that
for constructing $M^{({\rm tr})}_{3|{\rm loc}}(f,g,h)$,
one has to take the completely antisymmetric part
$M^{({\rm ad},a)}_{2|{\rm loc}}(z|g,h)$ of
$M^{({\rm ad})}_{2|{\rm loc}}(z|g,h)$ or, which is equivalent,
to take the antisymmetric part of the form
\begin{eqnarray*}
V(f,g,h)=\sigma(f,M^{({\rm ad})}_2)\int dzf(z)
M^{({\rm ad})}_{2|{\rm loc}}(z|g,h)=V_1(f,g,h)_1+V_2(f,g,h), \\
V_1(f,g,h)=a\int dzf(z)\left[g(z)\left(
\frac{\overleftarrow{\partial}}{\partial z^A}\omega^{AB}
\frac{\partial}{\partial z^B}\right)^3h(z)\right], \\
V_2(f,g,h)=\sigma(f,M_2)\int dzf(z)d_1^{\rm ad}T_1(z|g,h),\quad
T_1(z|g)=\sum_{p=0}^Kt^{(C)_p}(z)(\partial^z_C)^pg(z).
\end{eqnarray*}

As has been already noted above, 
the
functional $V_1(f,g,h)$ is antisymmetric w.r.t. $f$, $g$, and $h$.
The functional $V_2(f,g,h)$ can be rewritten as
\begin{eqnarray*}
V_2(f,g,h)=\sigma(f,M_2)\!\!\int\!\!dzf(z)[\sigma(g,M_2)\{g(z),\!T_1(z|g)\}\!-
\sigma(g,h)\sigma(h,M_2)\{h(z),\!T_1(z|g)\}\!- \\
-T_1(z|\{g,h\})]=\tilde{T}(\{f,g\},h)-\sigma(g,h)\tilde{T}(\{f,h\},g)-
\tilde{T}(f,\{g,h\})= \\
=[T_-(\{f,g\},h)-\sigma(g,h)T_-(\{f,h\},g)-T_-(f,\{g,h\})]+ \\
+[T_+(\{f,g\},h)-\sigma(g,h)T_+(\{f,h\},g)-T_+(f,\{g,h\})]\equiv \\
\equiv V_{2-}(f,g,h)+V_{2+}(f,g,h),
\end{eqnarray*}
where
\begin{eqnarray*}
\tilde{T}(f,g)=\int dz\sum_{p=0}^Kt^{(C)_p}(z)
(-1)^{|\varepsilon_C|_{1,p}\varepsilon(f)}f(z)(\partial^z_C)^pg(z), \\
T_{\pm}(f,g)=\frac{1}{2}[\tilde{T}(f,g)\pm\sigma(f,g)\tilde{T}(g,f)=
\pm\sigma(f,g)T_{\pm}(g,f).
\end{eqnarray*}
Note that the functionals $T_{\pm}(f,g)$ have the same structure as the
functional $\tilde{T}(f,g)$:
\begin{eqnarray*}
T_{\pm}(f,g)=\int dz\sum_{p=0}^Kt_{\pm}^{(C)_p}(z)
(-1)^{|\varepsilon_C|_{1,p}\varepsilon(f)}f(z)(\partial^z_C)^pg(z), \\
t_{\pm}^{(C)_p}(z)=t^{(C)_p}(z)\pm\sum_{q=p}^KC_q^p(\partial^z_D)^{q-p}
t^{(D)_{q-p}(C)_p}(z)(-1)^{q+(|\varepsilon_C|_{1,p}+|\varepsilon_D|_{1,q-p}+
\varepsilon(M_2))|\varepsilon_D|_{1,q-p}}.
\end{eqnarray*}

Obviously, the functional
$$
V_{2-}(f,g,h)=d_2^{\rm tr}T_-(f,g,h)
$$
is completely antisymmetric. By construction, the functional $V_{2+}(f,g,h)$
is antisymmetric with respect to $g$ and $h$. Imposing the antisymmetry
w.r.t. $f$ and $g$ yields
\[
T_+(f,\{g,h\})+\sigma(f,g)T_+(g,\{f,h\})=0,
\]
\begin{equation}
\sum_{p=0}^K\biggl(t_+^{(C)_p}(z)(\partial^z_C)^p[\{g(z),h(z)\}]-
(-1)^{p+(\varepsilon(M_2)+|\varepsilon_C|_{1,p})|\varepsilon_C|_{1,p}}
\{(\partial^z_C)^p[t^{(C)_p}(z)g(z)],h(z)\}\biggr)=0. \label{A7.3}
\end{equation}

Let $K>0$. In this equation, we consider the term with the highest derivative
of the function $h(z)$ whose order is equal to $K+1$.
It enters only the first term of (\ref{A7.3}) and has the form
\[
t_+^{(C)_K}(z)(\partial^z_C)^p\frac{\partial}{\partial z^A}h(z)=0.
\]
This implies that
\[
t_+^{(C)_K}(z)=0.
\]
Analogously, we make sure that all coefficient functions
$t_+^{(C)_p}(z)$ with $p\ge1$ vanish.
As a result, equation (\ref{A7.3}) is reduced to
\[
\{t^{(C)_0}(z),h(z)\}=0\quad\Longrightarrow\quad t^{(C)_0}(z)=b={\rm const},
\quad \varepsilon(b)=\varepsilon(M_2)+n_-,
\]
which gives
\begin{eqnarray}
V_2(f,g,h)=b\int dz\{f(z),g(z)\}h(z)=b\int dzf(z)\{g(z),h(z)\}. \label{A7.3a}
\end{eqnarray}
Here, the functional $V_2(f,g,h)$
is completely antisymmetric and, therefore, is a cocycle in the trivial representation.

For $n_+=n_-$, this cocycle is trivial. Indeed, we have
$$
\int dz\{f(z),g(z)\}h(z)= d^{{\rm tr}}_1 \mu(f,g,h),
$$
where
$$
\mu(f,g)= -\frac{1}{4}\int \di z\, z^A\frac{\partial f(z)}{\partial z^A} g(z)=-\sigma(f,g)\mu(g,f).
$$
Let us show that for $n_+\ne n_-$, the cocycle~(\ref{A7.3a}) is nontrivial.
Indeed, suppose the relation
\begin{eqnarray}
b\int dz\{f(z),g(z)\}h(z)=d_2^{\rm tr}M_2^{\prime{\rm tr}}(f,g,h)=\nonumber\\
=M_2^{\prime{\rm tr}}(\{f,g\},h)-\sigma(g,h)M_2^{\prime{\rm tr}}(\{f,h\},g)-
M_2^{\prime{\rm tr}}(f,\{g,h\}). \label{A7.4}
\end{eqnarray}
is valid.
Choosing the functions $f$, $g$, and $h$ such that
\[
{\rm supp}(h)\bigcap\left[{\rm supp}(f)\bigcup{\rm supp}(g)\right]=\varnothing,
\]
we obtain
\[
\hat{M}_2^{\prime{\rm tr}}(\{f,g\},h)=0,
\]
As shown in the section~\ref{Htr2}, this equation implies that
\[
M_2(f,g)=c\bar{f}\bar{g}+M_{2|2}(f,g),
\]
where the expression for the local form $M_{2|2}(f,g)$ is given by (\ref{4.3}).
Equation (\ref{A7.4}) is reduced to
\begin{eqnarray}
b\int dz\{f(z),g(z)\}h(z)=M_{2|2}(\{f,g\},h)-\sigma(g,h)M_{2|2}(\{f,h\},g)-
M_{2|2}(f,\{g,h\}). \label{A7.5}
\end{eqnarray}
The total order of the derivatives in the left-hand side is equal to 2 and
repeating
the arguments of section~\ref{Htr2}, we conclude that
one should set $Q=1$ in the expression (\ref{4.3}) for $M_{2|2}$.
Equation (\ref{A7.5}) is reduced to
\begin{eqnarray*}
\{f(z),g(z)\}=2[\sigma(A,f)\{m_2^A,f\}\frac{\partial g}{\partial z^A}-
\sigma(f,g)\sigma(A,g)\{m^A_2,g\}\frac{\partial f}{\partial z^A}]-m\{f,g\}+\\
+\{m,f\}g-\sigma(f,g)\{m,g\}f.
\end{eqnarray*}
Further following the arguments of section~\ref{Htr2}, we obtain
\[
m(z)\frac{\overleftarrow{\partial}}{\partial z^A}=0\quad\Longrightarrow\quad
m(z)=m={\rm const},\quad \varepsilon(m)=n_-+\varepsilon(M_2),
\]
\begin{equation}
2m_2^A(z)\frac{\overleftarrow{\partial}}{\partial z^C}\omega^{CB}-
2\sigma(A,B)m_2^B\frac{\overleftarrow{\partial}}{\partial z^C}\omega^{CA}+
(m+b)\omega ^{AB}=0, \label{A7.6}
\end{equation}
\[
(4+n_+-n_-)m=0.
\]
The general solution of equation (\ref{A7.6}) is
\[
m_2^A(z)=-\frac{1}{4}(m+b)z^A+
\frac{1}{2}m_2(z)\frac{\overleftarrow{\partial}}{\partial z^B}\omega^{BA}.
\]
Hence it follows that
\begin{equation}
 m=\frac{n_--n_+}{4}(m+b). \label{A7.6a}
\end{equation}
Since $n_+\ne n_-$, it follows from~(\ref{A7.6a}) that equation (\ref{A7.5}) has a solution only for $b=0$.

The general solution of the cohomology equation for
$M^{({\rm tr})}_{3|{\rm loc}}$ thus has the form
\begin{eqnarray}
M^{({\rm tr})}_{3|{\rm loc}}(f,g,h)=
a\int dzf(z)\left[g(z)\left(
\frac{\overleftarrow{\partial}}{\partial z^A}\omega^{AB}
\frac{\partial}{\partial z^B}\right)^3h(z)\right]+ \nonumber\\
+b\int dz\{f(z),g(z)\}h(z)+d_2^{\rm tr}M_2^{\rm tr}(f,g,h).\label{A7.6b}
\end{eqnarray}
For $n_+\ne n_-$, the first two terms in this expression are 
independent nontrivial cocycles.
Indeed, suppose that
\begin{eqnarray}
a\int dzf(z)\left[g(z)\left(
\frac{\overleftarrow{\partial}}{\partial z^A}\omega^{AB}
\frac{\partial}{\partial z^B}\right)^3h(z)\right]+b\int dz\{f(z),g(z)\}h(z)=
\nonumber \\
=d_2^{\rm tr}M_2^{\rm tr}(f,g,h). \label{A7.7}
\end{eqnarray}
Equation (\ref{A7.7}) means that the form
$ag(z)\left(\frac{\overleftarrow{\partial}}{\partial z^A}\omega^{AB}
\frac{\partial}{\partial z^B}\right)^3h(z)$
is a trivial cocycle in the space of the forms taking values
in adjoint representation
(recall that the form
$\{g(z),h(z)\}$
is trivial cocycle in this space).
As it was proved above, this is possible only for $a=0$. 
But we have just seen that
in this case,
equation (\ref{A7.7}) has a solution only for $b=0$.

If $n_+=n_-$, 
then the first term in~(\ref{A7.6b}) is the only nontrivial cocycle.

\vskip 5mm

{\bf Acknowledgements.} The work was supported by the RFBR (grants
No.~02-01-00930 (I.T.), No.~02-02-17067 (S.K.), and No.~02-02-16946
(A.S.)), by INTAS (grants No.~03-51-6346 (A.S.) and No.~00-00-262
(I.T.)) and by the grant LSS-1578.2003.2.


\begin{thebibliography}{99}

\bibitem{1} F.~Bayen, M.~Flato, C.~Fronsdal, A.~Lichnerovich and
D.~Sternheimer, {\it Ann.Phys}, {\bf 111}, 61--110 (1978); {\it
Ann.Phys}, {\bf 111}, 111--151 (1978).

\bibitem{2} M.~V.~Karasev and V.~P.~Maslov, {\it Nonlinear Poisson brackets. Geometry
and Quantization} [in Russian], Nauka, Moscow (1991); M.~V.~Karasev
and V.~P.~Maslov, {\it Nonlinear Poisson brackets. Geometry and
Qantization}, AMS, Providence, RI (1993).

\bibitem{3} B.~Fedosov, {\it Deformation Quantization and Index
Theory}, Akademie, Berlin (1996).

\bibitem{4} {\it M.Kontsevich}, ``Deformation Quantization of Poisson
Manifolds I,'' q-alg/9709040 (1997).

\bibitem{Zh} V.~V.~Zharinov, {\it Theor.~Math.~Phys.}, {\bf 136}, 1049--1065 (2003).

\bibitem{Leites} D.~A.~Leites and I.~M.~Shchepochkina, {\it Theor.~Math.~Phys.}, {\bf
126}, 281--306 (2001).

\bibitem{Ty1} I.~V.~Tyutin, {\it Theor.~Math.~Phys.}, {\bf 127}, 619--631 (2001).

\bibitem{Ty2} I.~V.~Tyutin, {\it Theor.~Math.~Phys.}, {\bf 128}, 1271--1292 (2001).

\bibitem{n=2} S.~E.~Konstein and I.~V.~Tyutin, ``Cohomologies of the
Poisson superalgebra on $(2,n)$-superdimensional spaces,''
hep-th/0411235.

\bibitem{central} S.~E.~Konstein and I.~V.~Tyutin,
``Deformations of the central extension of the
Poisson superalgebra'', hep-th/0501027.

\bibitem{Schei97} M.~Scheunert and R.~B.~Zhang, {\it J.Math.Phys.}, {\bf 39},
5024--5061 (1998), q-alg/9701037.

\bibitem{Ty3} S.~E.~Konstein, A.~G.~Smirnov and I.~V.~Tyutin, 
``General form of deformation of Poisson
superbracket,'' hep-th/0401023.

\bibitem{Hoermander} L.~H\"{o}rmander, {\it The Analysis of Linear Partial
Differential Operators}, Vol. I, Springer-Verlag,
Berlin--Heidelberg--New York--Tokyo (1983).


\end{thebibliography}
\end{document}